\documentclass[twocolumn,floatfix,amsmath,amssymb]{revtex4}

\usepackage{graphicx}
\usepackage{dcolumn}
\usepackage{color}
\def\ADD1#1{{\textcolor{blue}{#1}}}        
\def\ADD2#1{{\textcolor{red}{#1}}}        
  		
\usepackage{amssymb}

\newcommand{\be}{\begin{equation}}
\newcommand{\ee}{\end{equation}}

\begin{document}
\title{Tridimensional to bidimensional transition in magnetohydrodynamic turbulence with a guide field and kinetic helicity injection}

\author{N.E. Sujovolsky and P.D. Mininni}
\affiliation{Departamento de F\'{\i}sica, Facultad de Ciencias Exactas y 
                  Naturales, Universidad de Buenos Aires and IFIBA, CONICET, 
                  Buenos Aires 1428, Argentina }
\date{\today}

\begin{abstract}
We study the transition in dimensionality of a three-dimensional magnetohydrodynamic flow forced only mechanically when the strength of a magnetic guide field is gradually increased. We use numerical simulations to consider cases in which the mechanical forcing injects (or not) helicity in the flow. As the guide field is increased, the strength of the magnetic field fluctuations decrease as a power law of the guide field intensity. We show that for strong enough guide fields the helical magnetohydrodynamic flow can become almost two-dimensional. In this case, the mechanical energy can undergo a process compatible with an inverse cascade, being transferred preferentially towards scales larger than the forcing scale. The presence of helicity changes the spectral scaling of the small magnetic field fluctuations, and affects the statistics of the velocity field and of the velocity gradients. Moreover, at small scales the dynamics of the flow becomes dominated by a direct cascade of helicity, which can be used to derive scaling laws for the velocity field.
\end{abstract}
\maketitle

\section{Introduction}

Magnetohydrodynamic (MHD) turbulence is known to come in different flavors. Different regimes and scaling laws were reported in MHD flows depending on initial conditions \cite{ting1986turbulent,mininni2005numerical,lee2010lack,dallas2013symmetry,dallas2015self} or on how the system is forced \cite{dmitruk2003energy,zhou2004colloquium,perez2009role,grappin2010scaling,beresnyak2010scaling}. In recent years, the importance of anisotropy in these flows was discussed by several authors, specially in the context of the solar wind which at the largest scales can be modeled as an MHD flow with a magnetic guide field \cite{zhou2004colloquium}. In situ observations of the solar wind near Earth orbit and in the heliosphere show that the turbulence is dominated by fluctuations with wave vectors perpendicular to the guide field, i.e., that the flow has a strong two-dimensional (2D) component \cite{matthaeus_evidence_1990,dasso_anisotropy_2005}. Three-dimensional (3D) numerical simulations of MHD with a guide field and stirred magnetically also show a tendency of the system towards an approximately 2D MHD state \cite{ghosh_waves_1998,muller_statistical_2003}. A detailed numerical study using anisotropic forcing \cite{bigot_two-dimensional_2011} showed that the fraction of the energy in these 2D MHD modes increases as the amplitude of the guide field is augmented \cite{bigot_two-dimensional_2011}. This variety of regimes observed in MHD turbulence explains the lack of a clear phenomenological model for MHD flows at high Reynolds number, and whether a universal phenomenological theory can be developed is still an open question \cite{mininni_scale_2011}.

Other regimes of MHD turbulence were also reported in the literature. When an MHD fluid with a guide field has low conductivity (i.e., low magnetic Reynolds number), the system can suffer different transitions towards 2D regimes. Such transitions can result in two-dimensionalization and the suppression of turbulence \cite{moffat1967on}, or when stirred only mechanically, in a transition towards a 2D hydrodynamic (HD) regime \cite{knaepen_magnetohydrodynamic_2008}. In this limit, the flow rapidly suppresses magnetic field fluctuations perpendicular to the guide field as a result of Ohmic dissipation, making magnetic fluctuations negligible when compared to the external field. This is relevant particularly to  liquid metals. Laboratory experiments in the regime of low magnetic Reynolds number using gallium in a von K\'{a}rm\'{a}n flow confirmed that only small magnetic fluctuations are produced as the result of strongly anisotropic induction, and observed in some cases a power spectrum of magnetic fluctuations compatible with a $k^{-1}$ power law \cite{bourgoin_magnetohydrodynamics_2002}.

Recently, another regime of MHD turbulence displaying a transition towards a 2D HD state was discovered. In numerical simulations at high magnetic Reynolds number of 2D MHD flows and of 3D MHD flows with a guide field it was found that a transition towards a HD regime takes place when the ratio of mechanical to magnetic forcing exceeds a certain threshold, with the threshold depending on the scale at which the forcing is applied, on the anisotropy of the flow, and on the amplitude of the guide field in the 3D case \cite{alexakis_two-dimensional_2011,seshasayanan_edge_2014, seshasayanan_critical_2016}. The transition to the HD regime was accompanied by the development of an inverse cascade of energy, in which the system transfers a fraction of its energy from the injection scale to the largest scale available in the system, resulting in the growth of eddies with the size of the domain. For the 2D MHD case, the authors also showed that the transition to the HD regime is equivalent to a phase transition with the system behaving near the threshold as in the vicinity of a critical point, and that the behavior can be generic for other systems displaying inverse cascades after a transition \cite{seshasayanan_edge_2014, seshasayanan_critical_2016}.

The development of strong anisotropies with a transition from 3D to a 2D or quasi-2D regime is known to take place not only in MHD with a strong guide field \cite{moffat1967on,nazarenko_2d_2007,knaepen_magnetohydrodynamic_2008,alexakis_two-dimensional_2011,davidson_turbulence_2013} but in other systems as well, such as, e.g., HD turbulence with strong rotation \cite{pouquet10,Sen12,davidson_turbulence_2013,Gallet15}. In all these cases an external force imposes a preferred direction and is responsible for the departure of the flow from isotropy. Moreover, in many of these cases the accumulation of energy in 2D modes also results in the development of an inverse cascade of energy, as observed in \cite{seshasayanan_edge_2014, seshasayanan_critical_2016}. Also, if the system is dominated by the mechanical energy after the transition, in many cases the energy spectrum associated with the inverse cascade follows a $\sim k^{-5/3}$ power law, as observed for hydrodynamic turbulence in 2D \cite{paret_experimental_1997}.

The aim of the present work is to study 3D MHD turbulent flows with a strong guide field, forced only mechanically, and with large magnetic Reynolds number. In particular, we are interested in the transition of the system towards a 2D HD regime for sufficiently large values of the guide field. As the system is only stirred mechanically, magnetic fluctuations arise as the result of an induction process: for sufficiently large magnetic Reynolds number, the motion of the fluid elements can deform the guide field exciting small scale magnetic field fluctuations and MHD turbulence. However, as the amplitude of the guide field is increased, the magnetic field becomes more rigid and harder to deform, and magnetic field fluctuations decrease. As reported in \cite{alexakis_two-dimensional_2011}, for large guide fields this results in a regime in which only velocity field fluctuations are present, perpendicular to the guide field, and mostly 2D. Here, we extend the study in \cite{alexakis_two-dimensional_2011} to consider the case in which the mechanical forcing injects helicity in the flow.

The mechanical (or kinetic) helicity is a pseudo-scalar defined as
\begin{equation}
H=\int {\bf v} \cdot {\boldsymbol \omega} \; {\textrm d}V,
\label{eq:helicity}
\end{equation}
where ${\bf v}$ is the fluid velocity field and ${\boldsymbol \omega} = {\boldsymbol \nabla}\times{\bf v}$ is the vorticity. In ideal barotropic hydrodynamic flows, $H$ is conserved (but it is not conserved in MHD). In general, $H$ measures the number of links in the vortex lines, and the departure of the flow from mirror symmetry  \cite{moffatt_degree_1969}. Although mechanical helicity is not conserved in ideal MHD, it still plays an important role in this case \cite{pouquet1976strong,brandenburg2005astrophysical}: it is known that helical flows favor the dynamo mechanism, a process by which kinetic energy is converted into magnetic energy to sustain the MHD flow.

We therefore use numerical simulations to explore the transition from a 3D MHD flow to a 2D HD regime in an MHD system with guide field and with helical mechanical forcing, and compare the transition with the non-helical case. We show that the helical MHD flow still goes through the transition for large enough guide fields, and also behaves in a way reminiscent of the inverse cascade of mechanical energy observed in 2D HD turbulence. However, the presence of helicity changes the spectral scaling of the small magnetic field fluctuations, and affects the statistics of the velocity field at small scales as well as of the velocity gradients. Moreover, recent studies in HD flows indicate that when the energy suffers an inverse cascade, kinetic helicity can go through a direct cascade in which it dominates the direct flux and the scaling laws observed in the spectra at small scales; this was observed in rotating flows \cite{pouquet10,mininni_rotating_2010}, and in truncated versions of the Navier-Stokes equation \cite{Biferale13}. We show that the same behavior is observed in our system, with the direct flux of helicity dominating over the direct flux of energy.

\section{Numerical Simulations}

\begin{table}
\centering
\begin{tabular}{p{0.8cm} p{0.8cm} p{0.8cm} p{1.5cm} p{1.5cm} p{1.5cm} p{0.8cm}}
\hline \hline
Run & $\alpha_{h}$ & $|{\bf B}_{0}|$ & $\left< |{\bf v}|^{2} \right>^{1/2}_{t}$ & $\left< |{\bf b}|^{2} \right>^{1/2}_{t}$ & $R_{e}=R_{m}$ & $k_{\nu}$\\
\hline \hline
A0 & $0$ & $0$ & $1.2$ & $0$ & $1050$ & $170$\\
A2 & $0$ & $2$ & $1.3$ & $0.30$ & $1090$ & $130$\\
A4 & $0$ & $4$ & $2.2$ & $0.14$ & $1800$ & $110$\\
A8 & $0$ & $8$ & $2.5$ & $0.02$ & $2080$ & $90$\\
\hline
B2 & $\pi / 4$ & $2$ & $1.4$ & $0.31$ & $1200$ & $130$\\
B4 & $\pi / 4$ & $4$ & $1.8$ & $0.16$ & $1500$ & $110$\\

B8 & $\pi / 4$ & $8$ & $2.9$ & $0.03$ & $2380$ & $100$ \\
\hline \hline
\end{tabular}
\caption{Parameters for all runs: $ \alpha_{h} $ controls the kinetic helicity injection in the fluid ($ \pi / 4 $ corresponds to the maximum possible injection rate), $ | { \bf B }_{ 0 } | $ is the guide magnetic field amplitude , $ \left< | { \bf v } | ^{ 2 } \right> ^{ 1/2 }_{ t } $ and $ \left< | { \bf b } |^{ 2 } \right> ^{ 1/2 }_{ t } $ are the averaged in time r.m.s.~field fluctuations in the turbulent steady state of each run, $ R_{e } $ and $ R_{ m } $ are respectively the kinetic and magnetic Reynolds numbers, and $ k_{ \nu } $ is the Kolmogorov dissipation wavenumber.}
\label{tab:datos_simulaciones}
\end{table}

We solve numerically the MHD equations for an incompressible conducting fluid interacting with a magnetic field
\begin{equation}
\label{eq:n-s_mhd}
\dfrac{\partial {\bf v}}{\partial t}+{\bf v}\cdot{\bf \nabla}{\bf v}=-{\bf \nabla}(p+p_{m})+{\bf B}\cdot{\bf \nabla}{\bf b}+\nu \nabla^{2}{\bf v}+{\bf f},
\end{equation}
\begin{equation}
\label{eq:B_mhd}
\dfrac{\partial{\bf b}}{\partial t} + {\bf v}\cdot{\bf \nabla} {\bf b}={\bf B}\cdot{\bf \nabla} {\bf v} + \eta \nabla^{2}{\bf b},
\end{equation}
\begin{equation}
\label{eq:v_incomp}
{\bf \nabla}\cdot{\bf v}=0,
\end{equation}
\begin{equation}
\label{eq:B_incomp}
{\bf \nabla}\cdot{\bf b}=0,
\end{equation}
where ${\bf B}={\bf B}_{0}+{\bf b}$ with ${\bf B}_{0}$ an externally imposed guide field and ${\bf b}$ the magnetic field fluctuations, ${\bf v}$ the velocity field, $p_{m}=B^{2}/2$ is the magnetic pressure (with uniform mass density $\rho =1$), $\nu$ is the kinematic viscosity, $\eta$ the magnetic diffusivity, and ${\bf f}$ a mechanical forcing. Both fields are solenoidal as it follows from Eqs.~(\ref{eq:v_incomp}) and (\ref{eq:B_incomp}). The magnetic field is written in Alfv\'{e}nic units, and all quantities in the equations are dimensionless. Equations (\ref{eq:n-s_mhd}) and (\ref{eq:B_mhd}) then have two control parameters: the Reynolds number $R_{e}=UL/\nu$, and the magnetic Reynolds number $R_{m}=UL/\eta$, where $U$ and $L$ are the characteristic velocity and length of the flow. Another dimensionless number of interest is the magnetic Prandtl number, $P_{m}=\nu/\eta$, which measures the ratio of viscous to magnetic diffusion. In all the cases we will consider, $P_{m}=1$ and $R_{e}=R_{m} \approx 10^{3}$.

The MHD equations were solved numerically inside a periodic cubic box of volume $(2\pi)^{3}$ using a dealiased pseudo-spectral method and a second order Runge-Kutta scheme to evolve in time \cite{mininni_nonlocal_2008,mininni2011hybrid}. All runs have a spatial resolution of $512^{3}$ regularly spaced grid points, unless otherwise stated. The flow was mechanically forced at $k_{f}=10$, using a randomly generated isotropic forcing, and with no electromotive force applied. The fluid was started from rest, and integrated for 10 large-scale turnover times in all cases. We performed two sets of runs. Runs in set A correspond to runs with no helicity injection, while runs in set B correspond to runs with maximal helicity injection (see Table \ref{tab:datos_simulaciones}). The viscosity, magnetic diffusivity, and amplitude of the forcing are kept the same in all the simulations. Therefore, in each set the only parameter changed from run to run is the amplitude of the guide field $B_0 = |{\bf B}_0|$. To control the rate of helicity injection in the two sets we used the method described in \cite{pouquet1978numerical}. Namely, we generate two independent and solenoidal random vector fields ${\bf c}$ and ${\bf d}$, which are normally distributed, and centered around $k=k_{f}$ in Fourier space. Then, the mechanical forcing in Fourier space is given by
$\hat{\bf f}_{\bf k} = \hat{\bf c}_{\bf k} \cos \alpha_{h} + \hat{\bf d}_{\bf k} \sin \alpha_{h} + i {\bf k} \times (\hat{\bf c}_{\bf k} \sin \alpha_{h} + \hat{\bf d}_{\bf k} \cos \alpha_{h})/k$,
where the hat denotes Fourier transformed, and $\alpha_{h}$ is a parameter. It is easy to verify that the helicity of the mechanical forcing ${\bf f}$ is then proportional to $\sin (2 \alpha_{h})$. Thus, simulations in set A correspond to $\alpha_{h}=0$, while simulations with maximal helicity injection in set B correspond to $\alpha_{h}=\pi/4$.

\begin{figure}
\centering
\includegraphics[width=8.3cm]{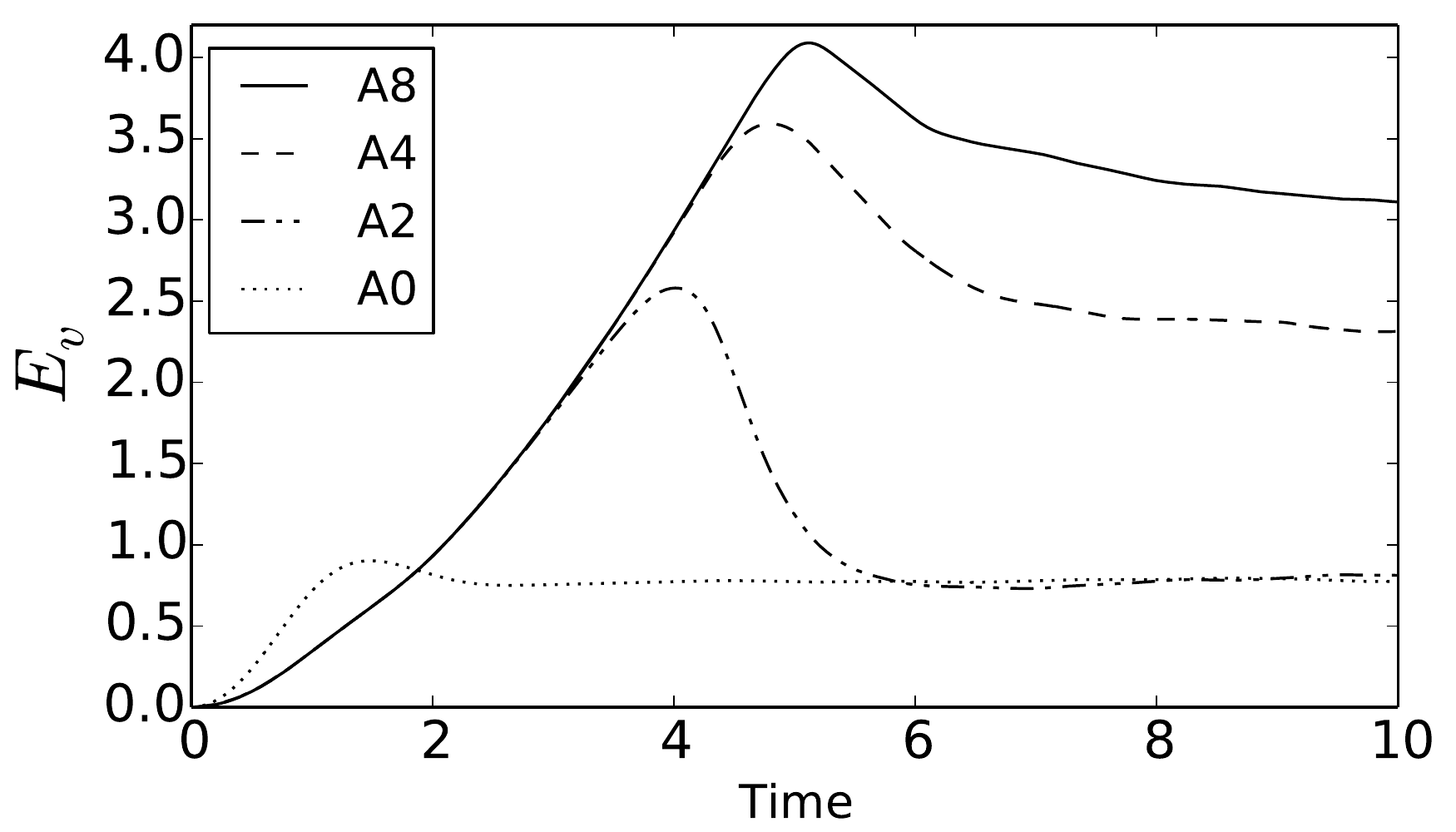} \\
\includegraphics[width=8.3cm]{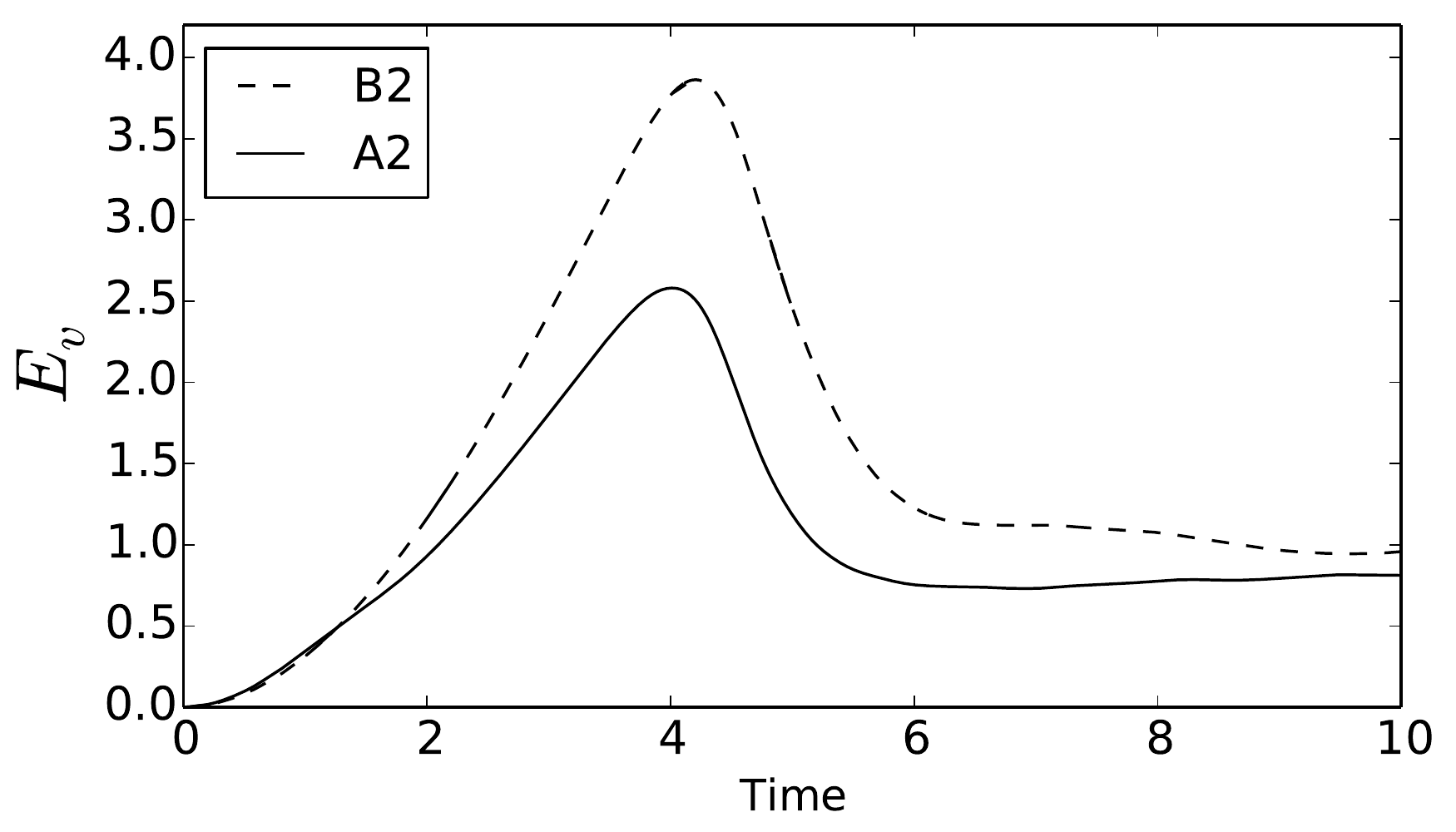}
\caption{{\it Top:} Kinetic energy as a function of time for simulations in set A, without helicity. The peak of kinetic energy increases with the strength of the guide field, with simulation A0 (dotted) having $B_0=0$, and simulation A8 (solid) having $B_0=8$. {\it Bottom:} Kinetic energy as a function of time for runs A2 (without helicity and with $B_{0}=2$) and B2 (with helicity, same $B_0$).}
\label{fig:E_vs_t}
\end{figure}

In the following we will need a way to quantify the anisotropy of the flow. This can be done by computing the energy spectrum in Fourier space, and energy fluxes. Considering the symmetry of the flows with the guide field, spectra can be computed isotropically, or in terms of parallel and perpendicular wave vectors (with respect to the direction of the guide field). As an example, for the isotropic kinetic energy spectrum, we have
\begin{equation}
\label{eq:integral}
E_v(k) = \frac{1}{2} \int \left| \hat{\bf v}({\bf k}') \right|^2 {\textrm d}S_{k},
\end{equation}
where $S_{k}$ is the surface on ${\bf k}'$ of the sphere of radius $k$ (in practice, in a discrete Fourier space the integral is replaced by a sum over all Fourier modes with $k \leq | {\bf k}' | < k + 1$). To define anisotropic spectra we can replace the surface of integration by a surface more appropriate to describe the flow anisotropy. Thus, the perpendicular kinetic energy spectrum $E(k_\perp)$ will be given by the sum over all Fourier modes with $k_\perp \leq | {\bf k}_\perp' | < k_\perp + 1$ (i.e., over cylindrical shells in Fourier space), where ${\bf k}_\perp$ is the projection of ${\bf k}'$ perpendicular to ${\bf B}_0$. In a similar way we can define the isotropic and perpendicular magnetic energy spectra $E_b(k)$ and $E_b(k_\perp)$, the helicity spectra $H(k)$ and $H(k_\perp)$, and perpendicular energy fluxes as described in more detail below.

\begin{figure}
\centering
\includegraphics[width=8.3cm]{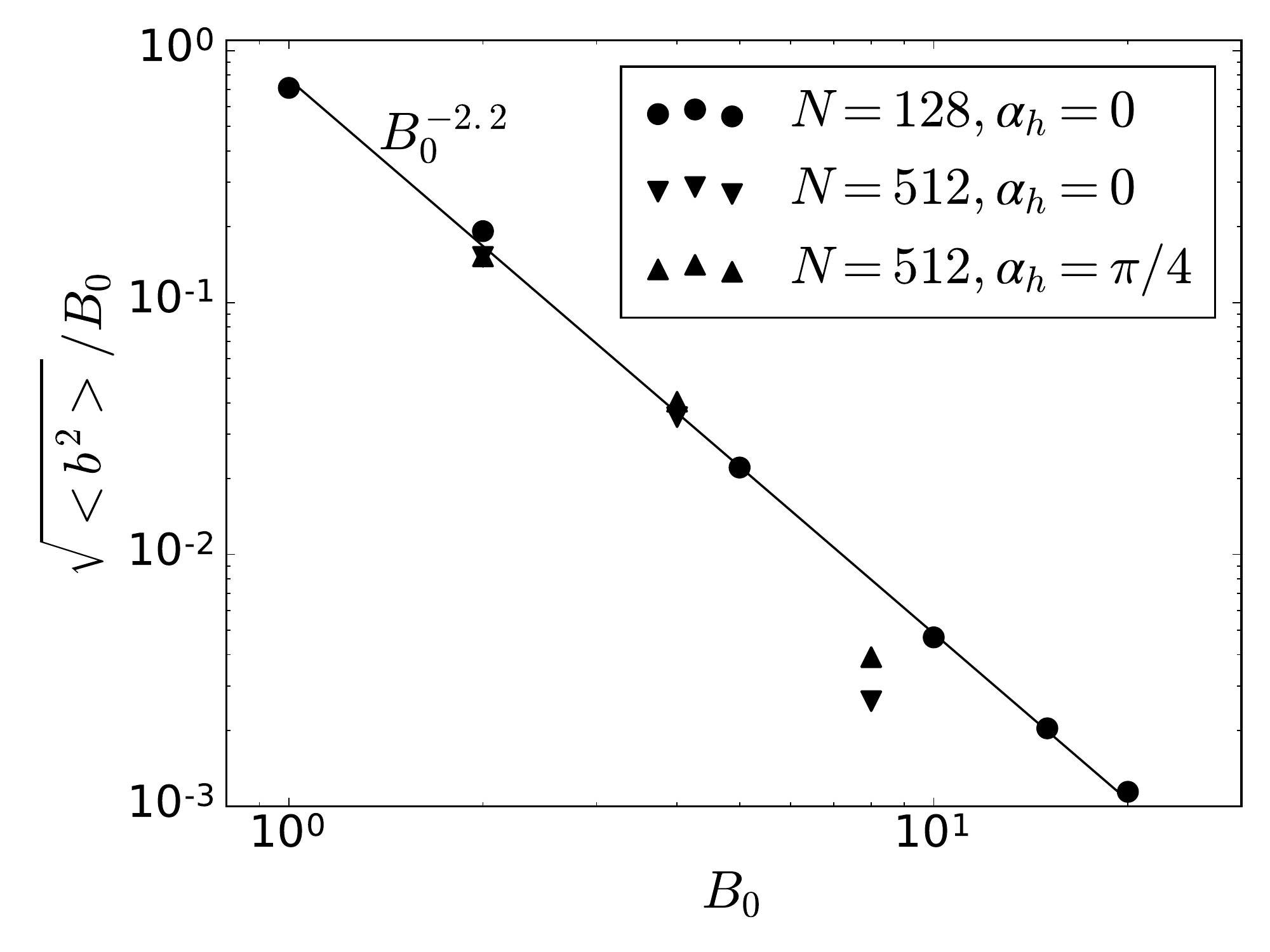}
\caption{Dimensionless ratio of r.m.s.~magnetic field fluctuations to the amplitude of the guide field, $(2E_{b})^{1/2}/B_{0}$, as a function of $B_0$, for the simulations in Table \ref{tab:datos_simulaciones} with $512^3$ grid points, and for several simulations with the same configuration but with spatial resolution of $128^3$ grid points. A power law resulting from a best fit to the data is shown as a reference.}
\label{fig:fluc_mag_b0}
\end{figure}

\begin{figure}
\centering
\includegraphics[width=8.3cm]{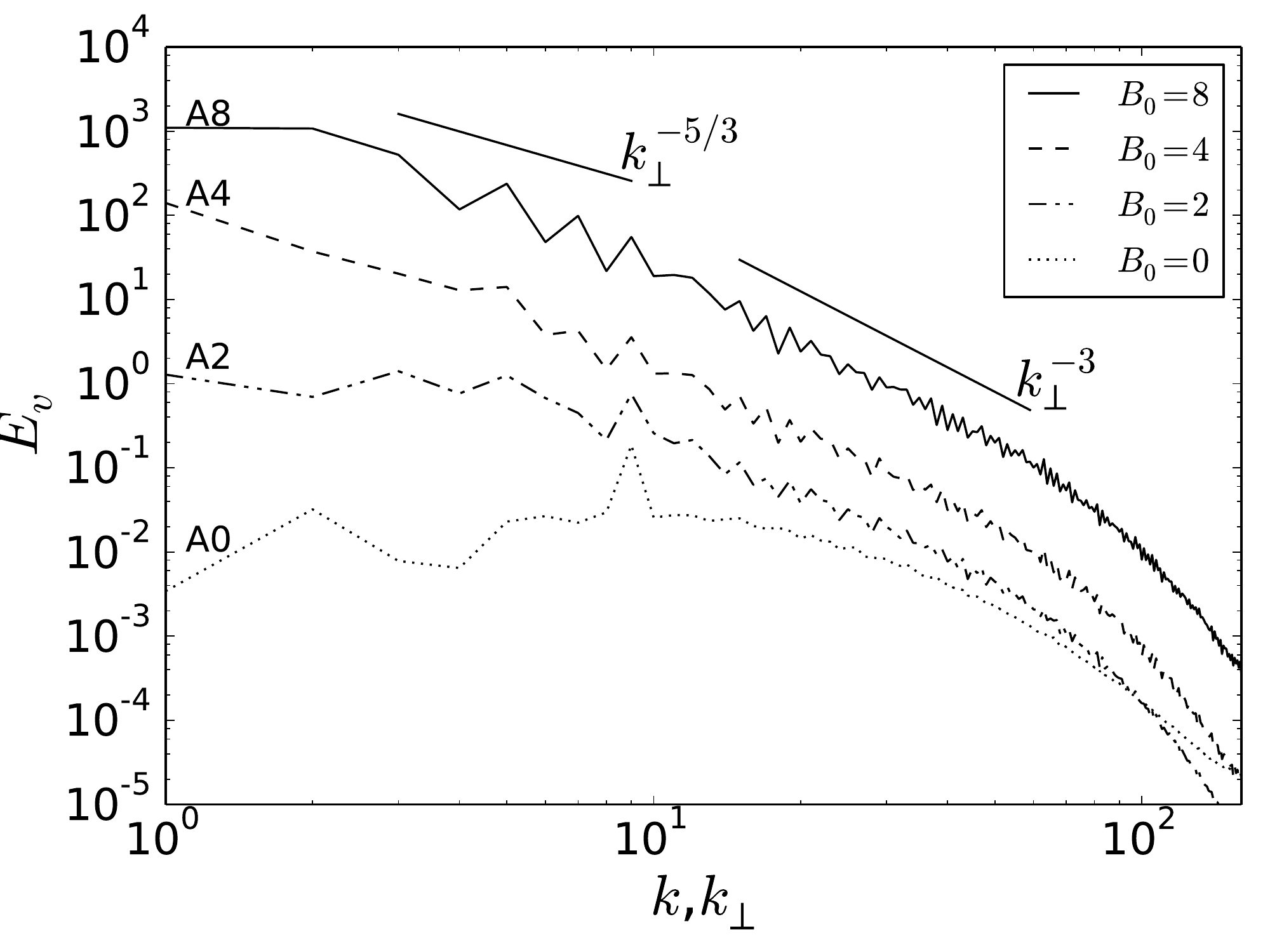}\\
\includegraphics[width=8.3cm]{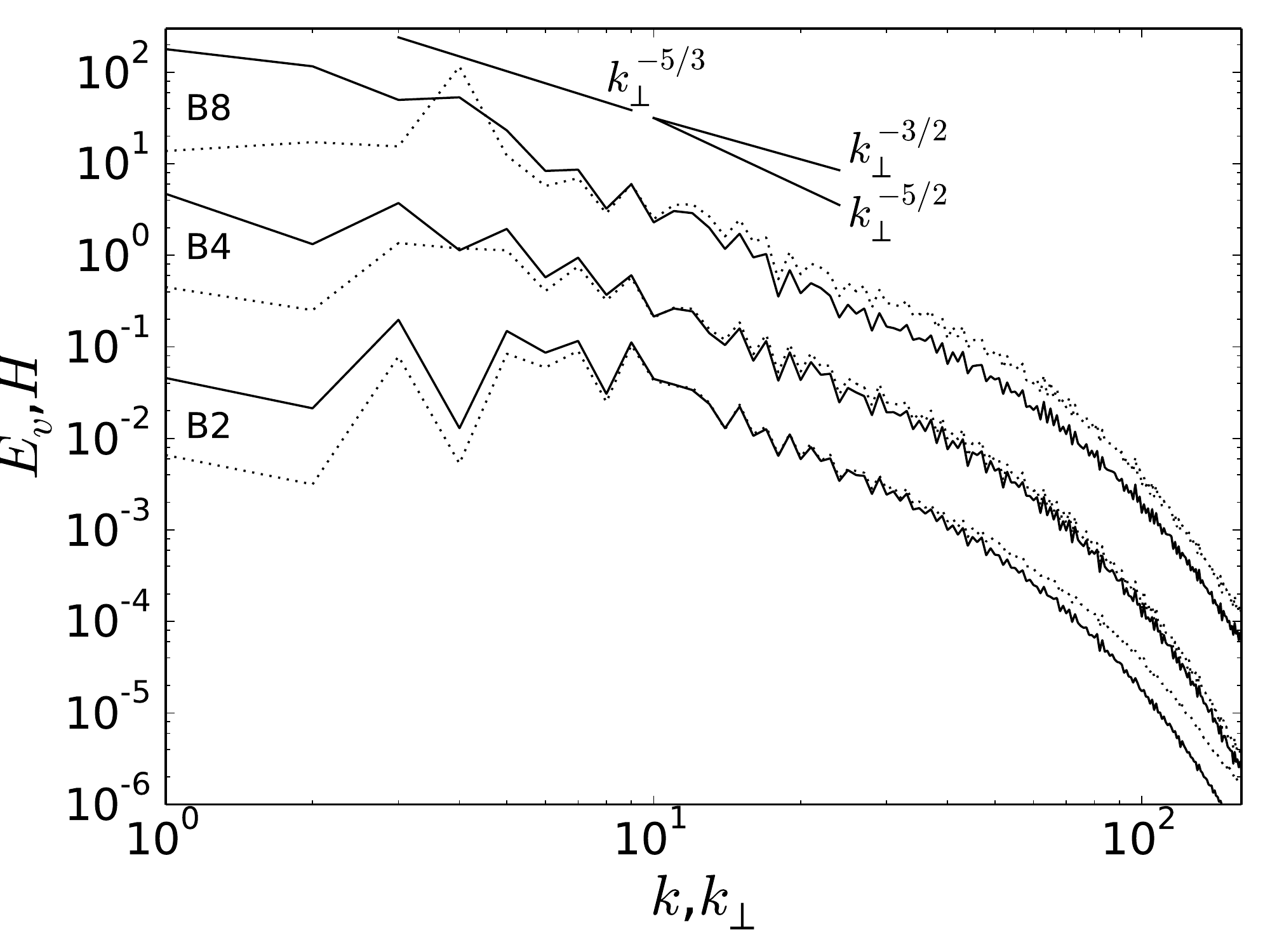}
\caption{{\it Top:} Kinetic energy spectrum for simulations in set A (without helicity). {\it Bottom:} Kinetic energy spectrum (solid) and helicity spectrum normalized by $k_{f}$ (dotted) for simulations in set B (with helicity). The spectra have been shifted vertically for better visualization, and the slopes indicate several power laws as references (see text for detail). The isotropic spectrum $E_v(k)$ is shown for runs A0, A2 and B2, in all other cases we show $E(k_\perp)$.}
\label{fig:Ev_vs_k}
\end{figure}

\section{Results}

\subsection{Kinetic and magnetic energy}

We start by discussing the general evolution of all simulations. Figure \ref{fig:E_vs_t} shows the time evolution of the kinetic energy for all simulations with non-helical mechanical forcing, and also compares the evolution of runs B2 and A2 (respectively with and without kinetic helicity injection). In all cases the kinetic energy grows monotonically until reaching a peak, which increases as the guide field is increased.
Note that as the fluid is started from rest, the flow must undergo an instability to generate turbulence. At early times, the kinetic energy increases as the result of the energy injected by the forcing, and dissipation remains slow (thus energy keeps accumulating in the system) until turbulence develops and the dissipation rate increases. The external magnetic field introduces a privileged axis and has a stabilizing effect in the flow, thus the system must reach larger values of the kinetic energy before becoming unstable. After this time (which also increases with $B_0$), the dissipation rate reaches a turbulent steady value, and the kinetic energy drops to also reach its saturation value. Interestingly, as $B_0$ increases, so does the kinetic energy in the turbulent regime.

\begin{figure}
\centering
\includegraphics[width=8.3cm]{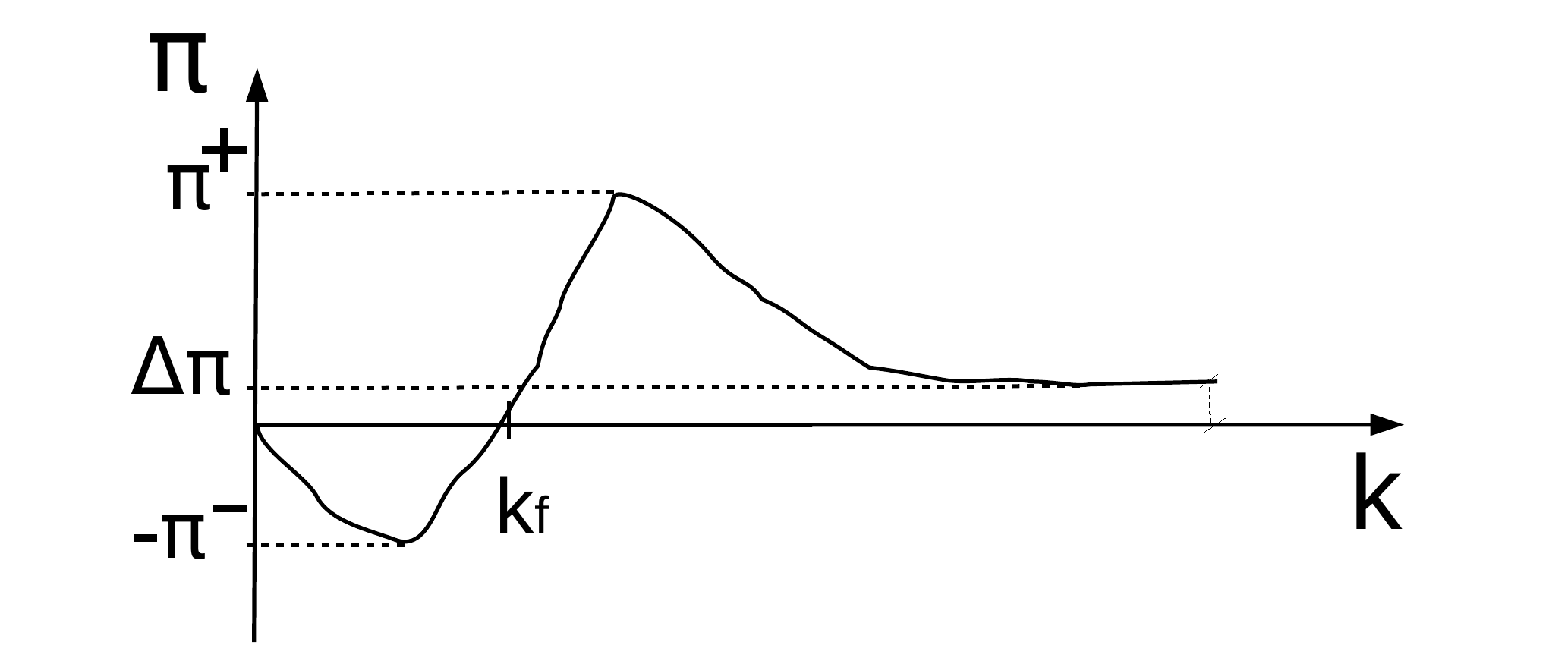}
\caption{Diagram showing a typical flux (either of kinetic energy, magnetic energy, or kinetic helicity) as a function of $k$ for simulations with ${\bf B}_{0} \neq 0$. In the diagram we show several characteristic values used for the analysis: the scale injection $k_{f}$ in which the flux changes sign, the maximum value of positive flux $\Pi^{+}$ (i.e., of flux towards small scales), the minimum value of negative flux $-\Pi^{-}$ (i.e., of inverse flux), and the value of the flux when $k \rightarrow k_{max}$, $\Delta \Pi$.} 
\label{fig:esquema_flujo}
\end{figure}

\begin{figure}
\centering
\includegraphics[width=8.3cm]{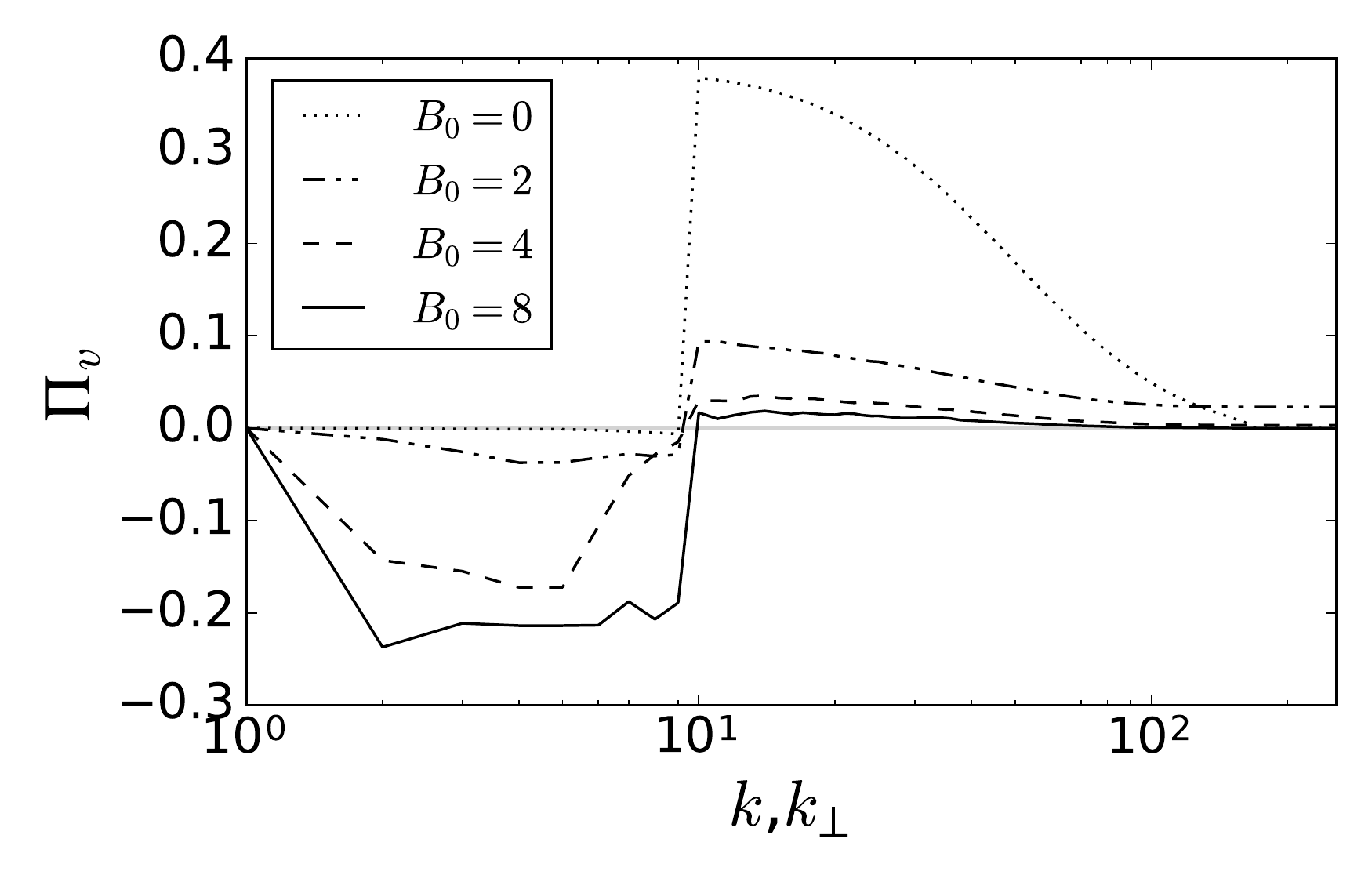}\\
\includegraphics[width=8.3cm]{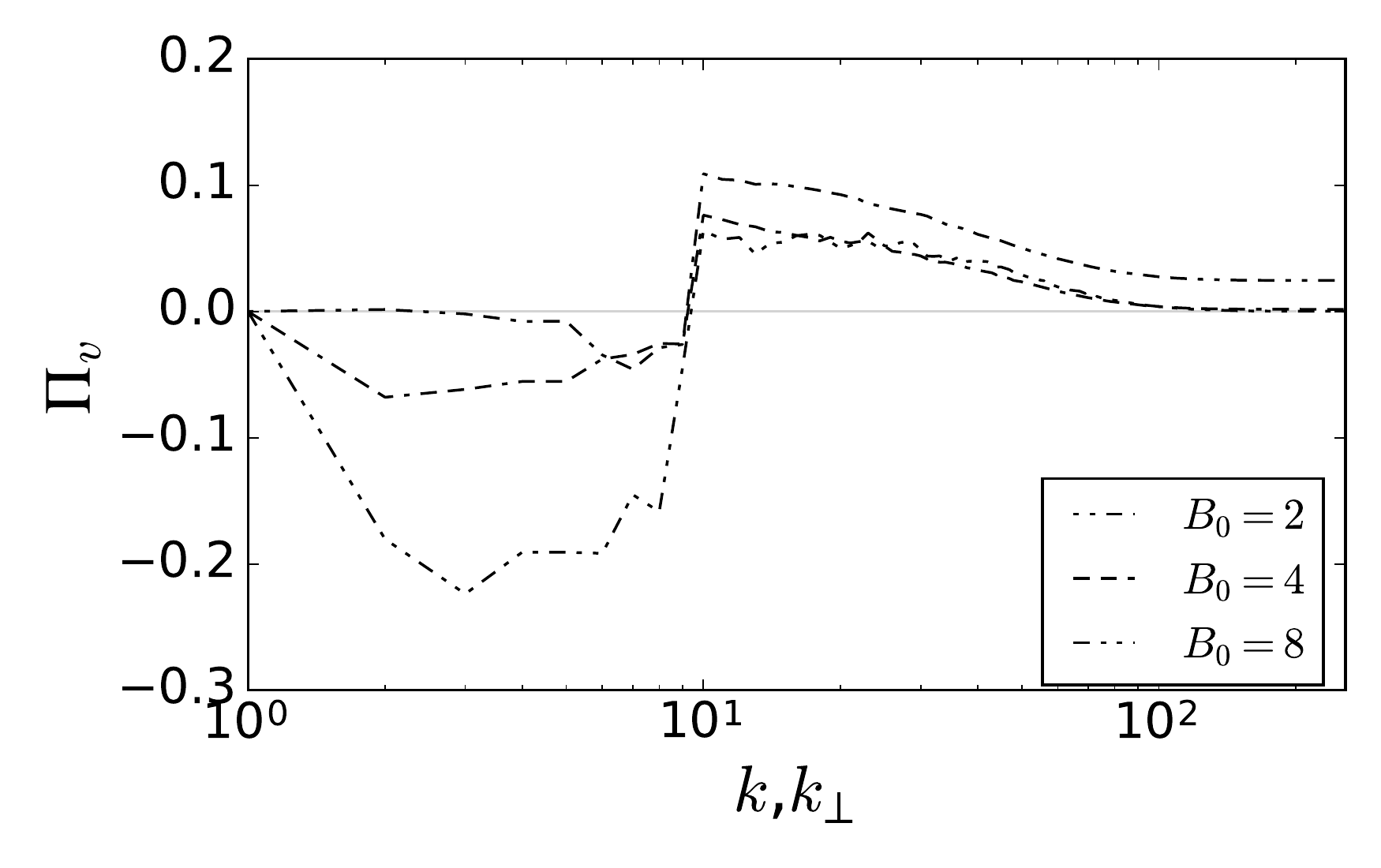}
\caption{{\it Top:} Flux of kinetic energy for runs in set A (without helicity), for different values of $B_0$, and time-averaged for long times on the inverse cascade scales. {\it Bottom:} Same for runs in set B (with helicity). The flux is shown as a function of $k$ for run A0, and as a function of $k_\perp$ for all other runs. In both cases, $\Pi^+$ decreases as $B_0$ is increased, negative values of the flux are observed for $k<k_f$ for large values of $B_0$, and $\Delta \Pi$ decreases towards zero.}
\label{fig:PIv_vs_k}
\end{figure}

The simulations with mechanical helicity behave similarly, but the maximum of energy (and the time to reach the maximum) also increases (see Fig.~\ref{fig:E_vs_t}). This is the effect of helicity, which also stabilizes the flow and slows down the instabilities. From Eq.~(\ref{eq:helicity}), a helical flow tends to have the velocity field parallel to the vorticity. The nonlinear term in the momentum equation can be rewritten as $ {\bf v} \times {\boldsymbol \omega} - {\bf \nabla} (p + {\bf v}^{2} / 2) $. Therefore, in a helical flow the term $ {\bf v} \times {\boldsymbol \omega} $ tends to be smaller, and larger velocities (or Reynolds numbers) are needed to destabilize the flow and transfer energy to scales different than the forced scale. After this happens, the flow rapidly evolves to a turbulent steady state.

The energy of magnetic fluctuations has a different fate. As the system is only forced mechanically, magnetic fluctuations grow as the result of the deformation of the guide field lines: for infinite $R_m$, the magnetic field lines are frozen to the flow. With finite (but still large) $R_m$, magnetic field lines are advected by the flow, and also diffuse by Ohmic dissipation. The advection of ${\bf B}_0$ by the turbulent flow creates small scale magnetic field fluctuations, which first grow in time, and then saturate to a steady r.m.s.~value in the turbulent regime. However, as $B_0$ increases, the guide field becomes more rigid, and energy in the magnetic field fluctuations decreases. Figure \ref{fig:fluc_mag_b0} shows the square root of the energy of magnetic fluctuations normalized by the amplitude of the guide field, $(2 E_{b})^{1/2}/B_{0}$, averaged at late times in the simulations, and as a function of $B_{0}$ for all runs. Besides the simulations with $512^{3}$ grid points, we also show the results for a large number of similar simulations using $128^{3}$ grid points. Overall, the data is compatible with a dependence $\left< b^2 \right>^{1/2} \sim B_0^{-2.2}$ independently of the helicity content of the flow, and where the exponent $-2.2$ was obtained from a best fit to the data. Note that for large values of $B_0$ energy in magnetic field fluctuations is negligible when compared to the kinetic energy. As in previous studies \cite{alexakis_two-dimensional_2011,seshasayanan_edge_2014,seshasayanan_critical_2016}, the system seems to undergo a transition towards a HD regime as $B_0$ is increased, with $|{\bf  B}_{0}|$ acting as the order parameter of the transition. Below we consider energy spectra and fluxes to show that for large $B_0$ the flow also approaches a quasi-2D state.

\begin{figure}
\centering
\includegraphics[width=8.3cm]{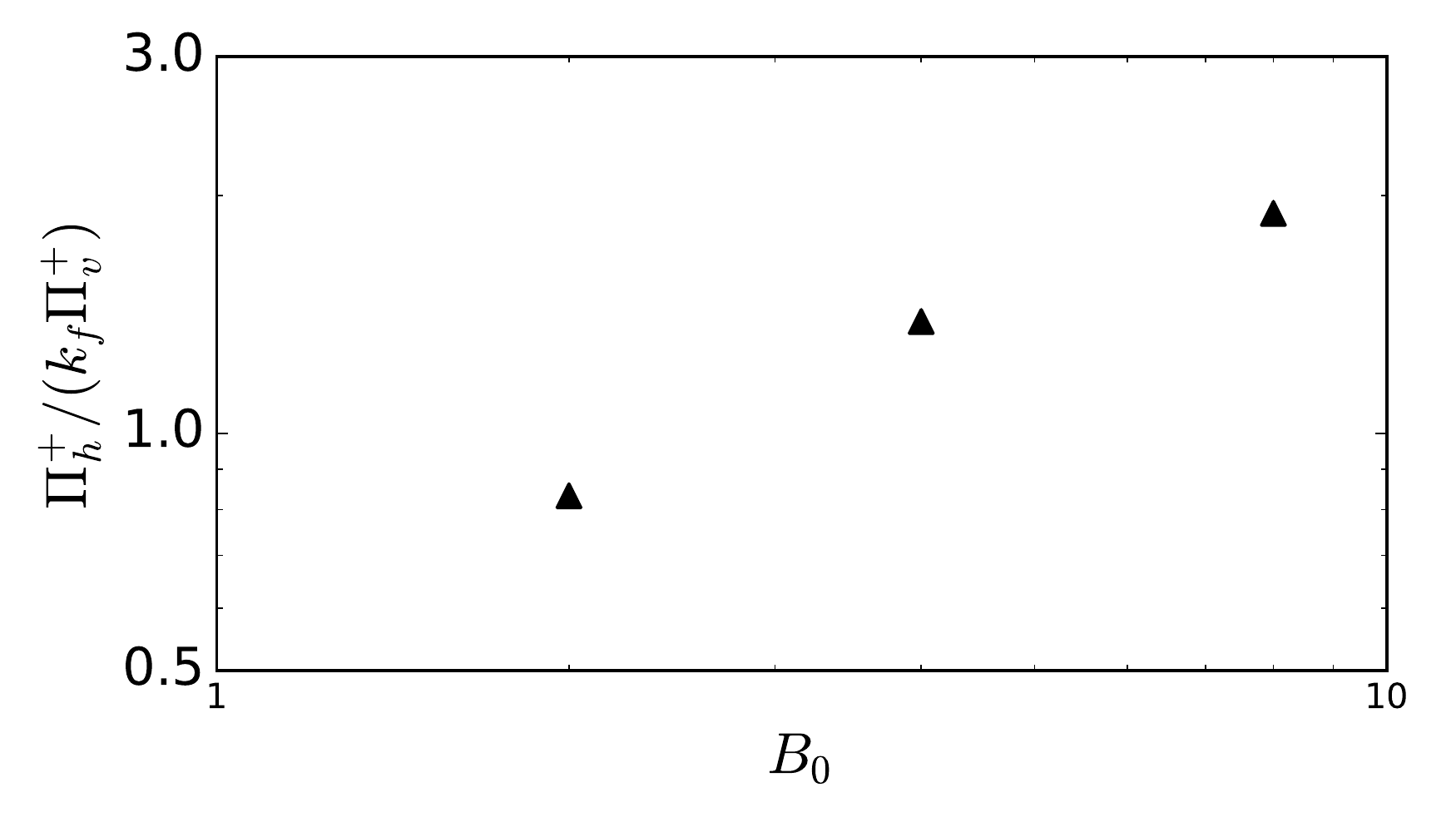}
\caption{Ratio of maximum direct helicity flux $\Pi_h^+$ to the maximum of direct kinetic energy flux $\Pi_v^+$ normalized by $k_{f}$, as a function of $B_{0}$.}
\label{fig:PIh_sobre_PIv}
\end{figure}

\subsection{Kinetic energy spectrum}

For late times in all simulations we computed the (temporal averaged) kinetic energy spectrum as a function of $k$ (for run A0 which is isotropic, and for runs with $B_{0}=2$ which are weakly anisotropic), and as a function of $k_\perp$ (for runs with $B_{0}=4$ and $8$, which are anisotropic). All spectra are shown in Fig.~\ref{fig:Ev_vs_k}. The simulation without a guide field (A0) results in just a hydrodynamic turbulent flow, as there are no sources of magnetic field fluctuations. In this run, a direct cascade of energy is observed, with a short inertial range compatible with a Kolmogorov power law $\sim k^{-5/3}$ for wave numbers $k>k_{f}=10$ where energy is injected by the forcing (note that the scale separation used between the forcing scale and the box size, to allow for an inverse cascade if needed, reduces the range of scales available for a direct cascade inertial range). For $k \lesssim 10$ there are no significant energy excitations, nor a clear scaling in the spectrum.

As the magnetic field is increased in Fig.~\ref{fig:Ev_vs_k} we observe two changes in the spectrum: On the one hand, we observe the appearance of an inverse transfer of kinetic energy, with the energy spectrum peaking at small values of $k_\perp$. This is particularly evident for runs A4 and A8 (respectively, with $B_{0}=4$ and $8$). As a reference, we show in Fig.~\ref{fig:Ev_vs_k} for $k_\perp <10$ a $k_\perp^{-5/3}$ power law, which corresponds to the slope of the energy spectrum in the inverse cascade range of 2D HD turbulence (note that in these runs, magnetic field fluctuations are negligible and the system is almost in a hydrodynamic regime, see Fig.~\ref{fig:fluc_mag_b0}). On the other hand, we observe the appearance of a much steeper spectrum in a broad range of wave numbers with $k_\perp>10$. All simulations with non-helical forcing (runs A) and large guide field show a spectrum compatible with a power law $\sim k_\perp^{-3}$, which is the spectrum of energy in the direct cascade range of 2D HD turbulence \cite{batchelor1969computation}.

The simulations with kinetic helicity injection (see Fig.~\ref{fig:Ev_vs_k}) also show a change in the kinetic energy spectrum for large $B_0$, but with certain differences with respect to the simulations in set A. A pile up of energy at small wave numbers is still observed (the $k_\perp^{-5/3}$ power law is also shown as a reference), and the spectrum at large wave numbers also becomes steeper than in the case of isotropic MHD (and HD) turbulence. However, the slope of the kinetic energy spectrum for $k_\perp>10$ seems to be less steep than in the simulations without helicity. In Fig.~\ref{fig:Ev_vs_k} we show a power law $\sim k_\perp^{-5/2}$ only as a reference, we will come back to the slope of this spectrum later.

\begin{figure}
\centering
\includegraphics[width=8.3cm]{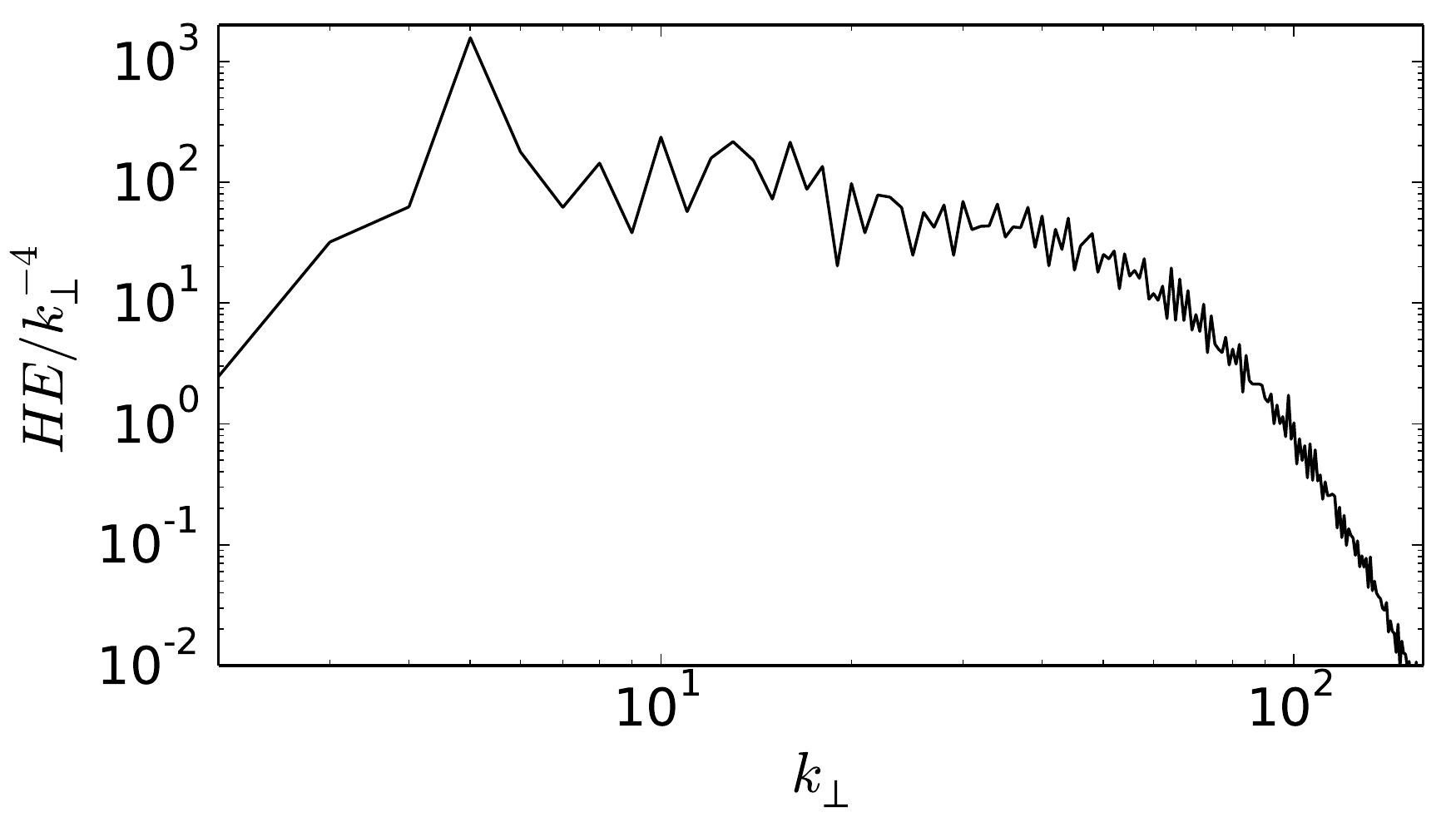}\\
\includegraphics[width=8.3cm]{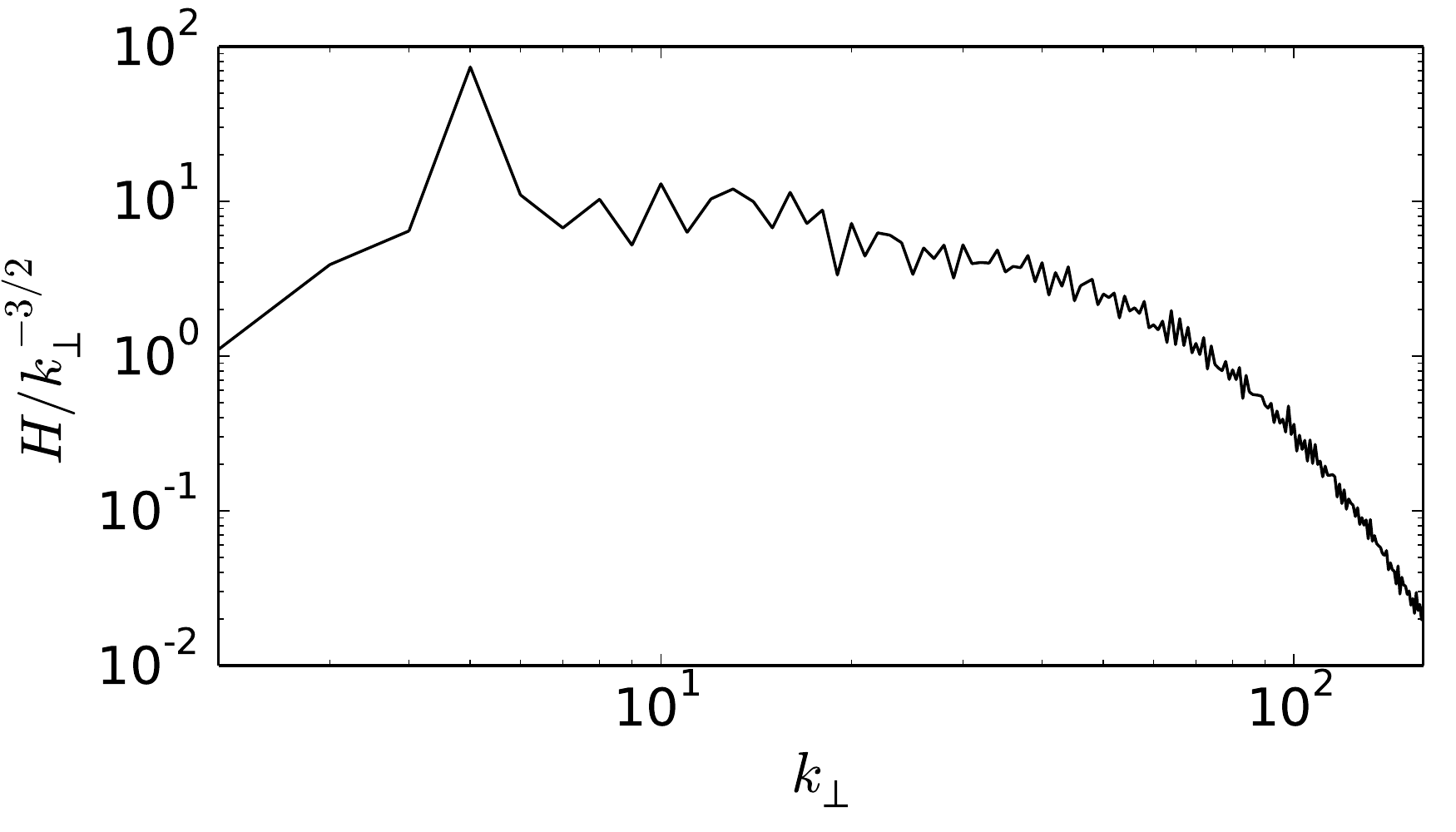}\\
\caption{{\it Top:} Compensated spectrum $H(k_{\bot}) E(k_{\bot}) / k_{\bot}^{-4}$ for run B8 (with mechanical helical forcing, and $B_0=8$). {\it Bottom:} Spectrum of helicity $H(k_\perp)$ in the same run, compensated by $k_\perp^{-3/2}$.}
\label{fig:E_H_compensados}
\end{figure}

\begin{figure}
\centering
\includegraphics[width=8.3cm]{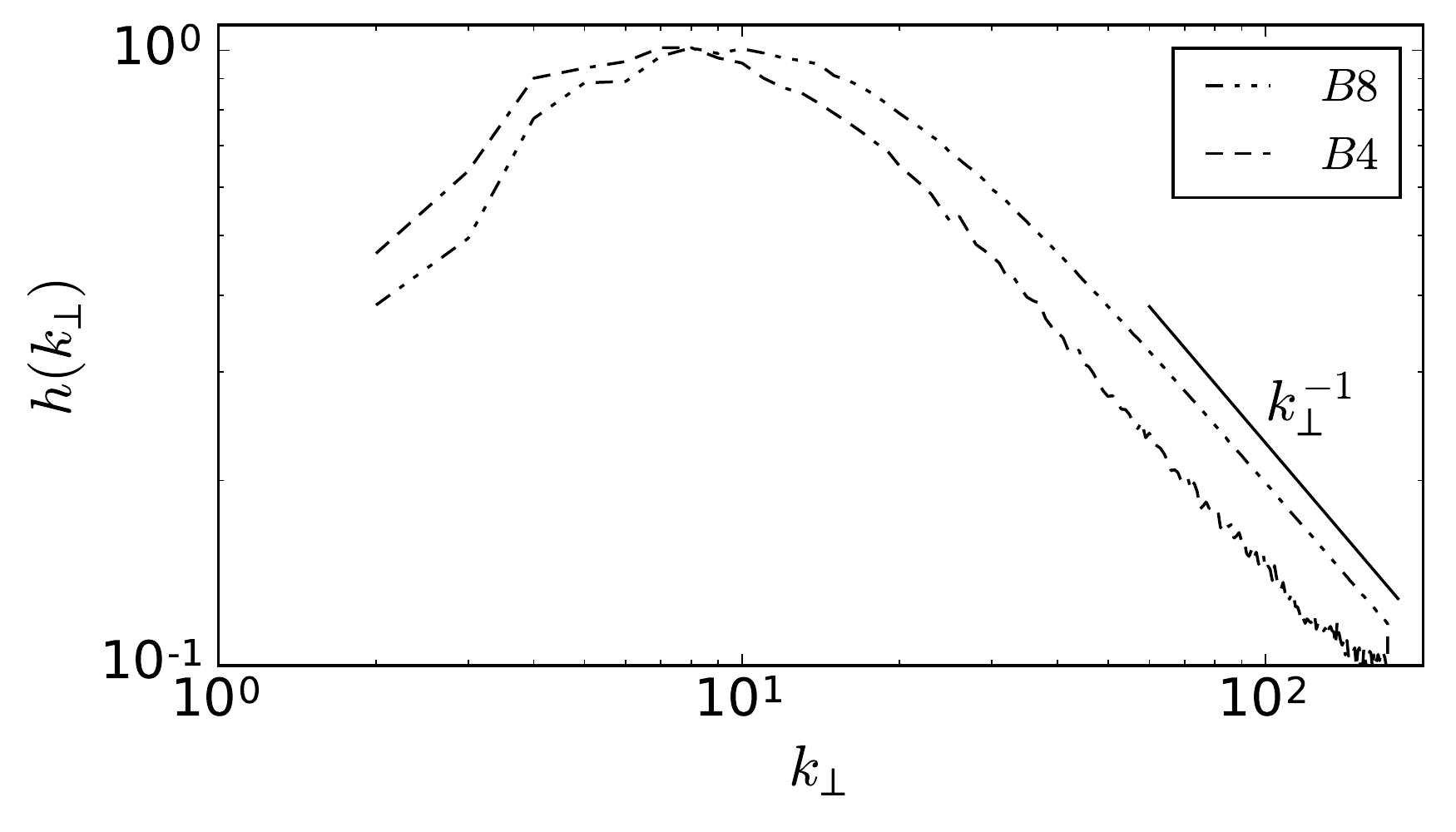}
\caption{Relative helicity spectrum $h(k_\perp)$ for runs B4 and B8 (both with kinetic helicity, and respectively with $B_0=4$ and $8$). A slope $k_\perp^{-1}$ is shown as a reference. Note the relative helicity spectrum is shallower than $k_{\bot}^{-1}$ everywhere except in the dissipative range.}
\label{fig:Hr_vs_k}
\end{figure}

For the simulations in set B we are also interested in the spectrum of kinetic helicity, which are also shown in Fig. \ref{fig:Ev_vs_k}. All helicity spectra show a transfer of helicity towards wave numbers larger than $k_f$, and as $B_0$ increases this range of the helicity spectrum becomes shallower than the energy spectrum (note the separation of the two spectra for $k>k_f$ in run B8). As a reference, we show a power law $\sim k_\perp^{-3/2}$ for this range, which is also discussed in detail below. Interestingly, there are also clear differences between the spectra $H(k)$ and $E_v(k)$ for wave numbers smaller than the forcing wavenumber. The spectrum of helicity does not peak at $k=1$ or $k_\perp=1$ even for large $B_0$, indicating there is no significant transfer of helicity towards small wave numbers. This is compatible with the fact that helicity cannot be transferred towards large scales in any flow with finite energy, as from Eq.~(\ref{eq:helicity}) and from Schwarz inequality, $H(k) \leq k E(k)$, which gives $H(k) \to 0$ for $k \to 0$ if the energy in the flow is finite.

\subsection{Energy and helicity fluxes}
\begin{figure}
\centering
\includegraphics[width=8.3cm]{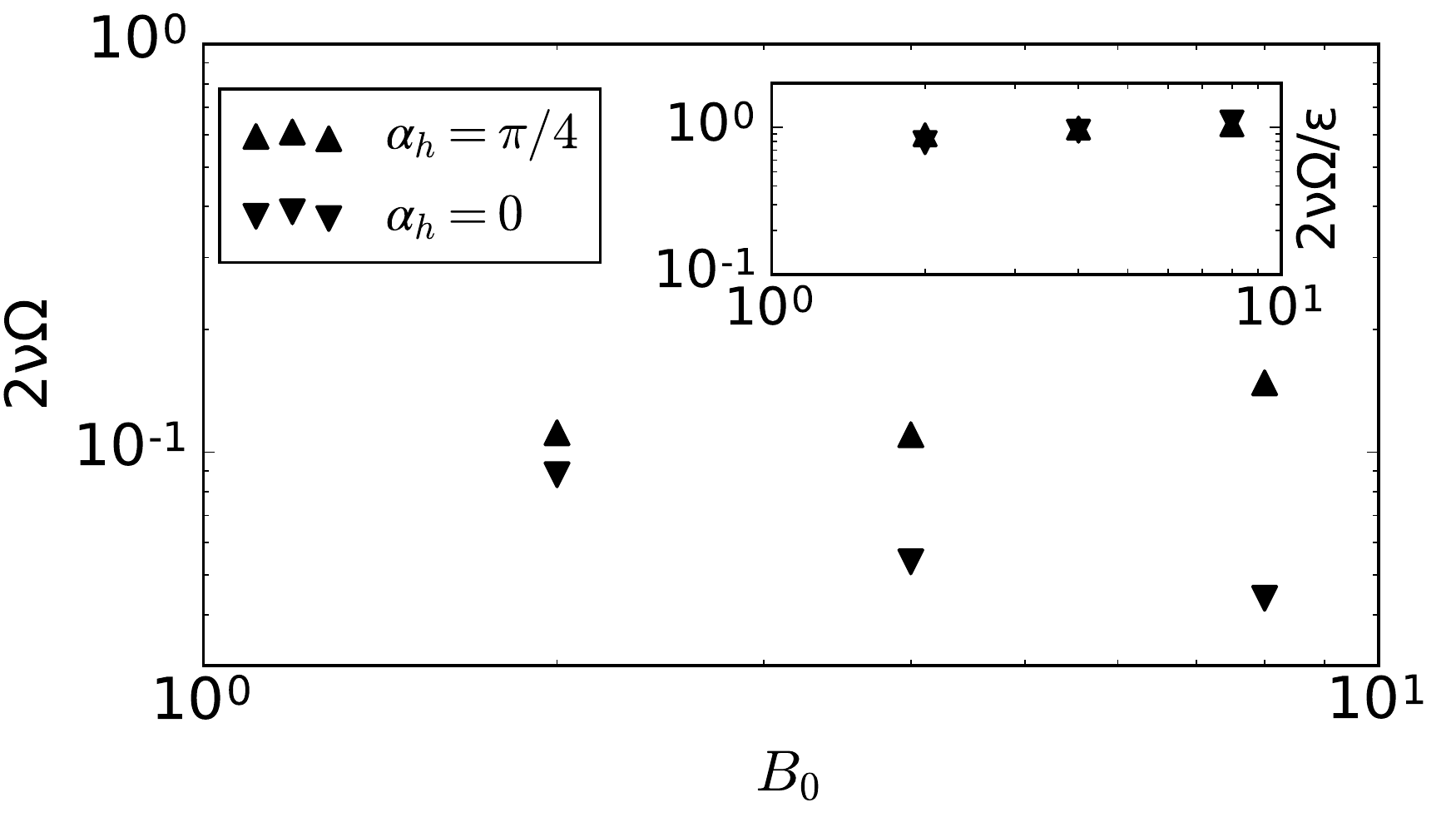}
\caption{Kinetic energy dissipation rate ($2 \nu \Omega$) as a function of $B_{0}$, for simulations with and without helical forcing. The inset shows the kinetic energy dissipation rate normalized by the energy injection rate ($\epsilon$) as a function of $B_{0}$. As expected for the case with negligible magnetic fluctuations, for large $B_{0}$ this ratio approaches unity, as energy can only be dissipated by velocity fluctuations.}
\label{fig:disip_b0}
\end{figure}
The results suggest that for large $B_0$ the system becomes almost hydrodynamic, it develops an inverse transfer of kinetic energy independently of the helicity content of the flow, and a direct transfer and cascade towards smaller scales that depends on whether the system has kinetic helicity or not. Confirmation of these results requires studying the flux of energy across scales. From Eq.~(\ref{eq:n-s_mhd}) the kinetic energy ``flux'' is obtained as
\begin{equation}
\begin{split}
\label{eq:PI_e_cin}
\Pi_{v}(k)=- \sum \limits_{k'=0}^{k} \int \left[ {\bf v}_{k'} \cdot \left( \widehat{ {\bf v} \cdot {\bf \nabla} {\bf v}} \right)_{k'} - \right. \\ \left. {\bf v}_{k'} \cdot \left( \widehat{ {\bf B} \cdot {\bf \nabla} {\bf b}}\right)_{k'} \right] {\textrm d}S_{k'},
\end{split}
\end{equation}
where the hat ($\, {\widehat \ } \, $) denotes the Fourier transform as before. From Eq.~(\ref{eq:B_mhd}) a ``flux'' of magnetic energy is obtained as
\begin{equation}
\label{eq:PI_e_mag}
\begin{split}
\Pi_{b}(k)=- \sum \limits_{k'=0}^{k} \int \left[ {\bf b}_{k'} \cdot \left( \widehat{ {\bf v} \cdot {\bf \nabla} {\bf b}} \right)_{k'} - \right. \\ \left. {\bf b}_{k'} \cdot \left( \widehat{ {\bf B} \cdot {\bf \nabla} {\bf v}} \right) _{k'} \right] {\textrm d}S_{k'}.
\end{split}
\end{equation}
Finally, we define the ``flux'' of kinetic helicity as usual using the hydrodynamic expression
\begin{equation}
 \begin{split}
\label{eq:PI_h_cin}
\Pi_{h}(k)=- \sum \limits_{k'=0}^{k} \int \left[ {\boldsymbol \omega}_{k'} \cdot \left( \widehat{ {\bf v} \cdot {\bf \nabla} {\bf v}} \right)_{k'} + \right. \\ \left. {\bf v}_{k'} \cdot {\bf \nabla} \times \widehat{ \left( {\bf v} \cdot {\bf \nabla}  {\bf v}\right)}_{k'} \right]  {\textrm d}S_{k'}.
\end{split}
\end{equation}

\begin{figure}
\centering
\includegraphics[width=8.3cm]{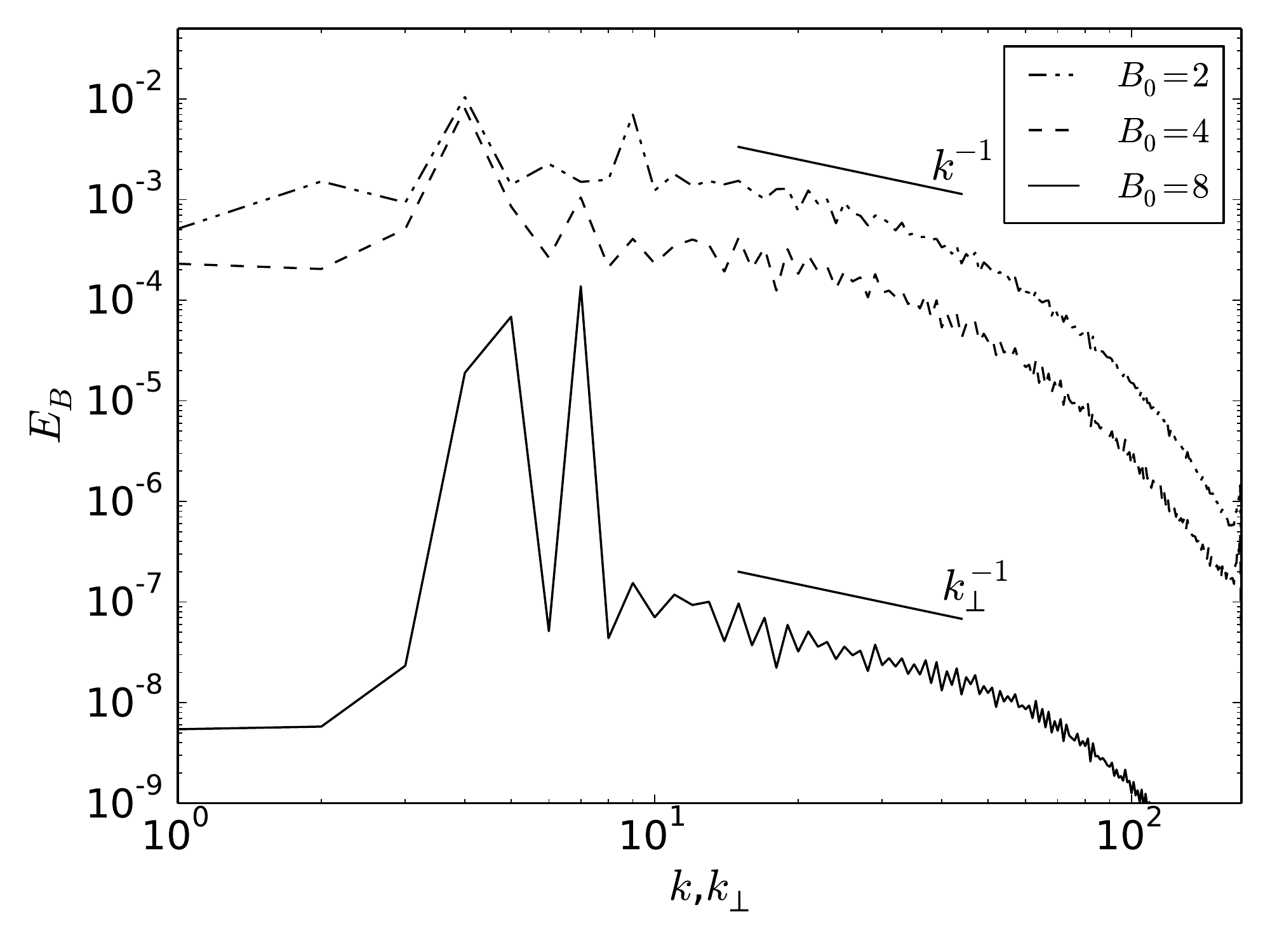}\\
\includegraphics[width=8.3cm]{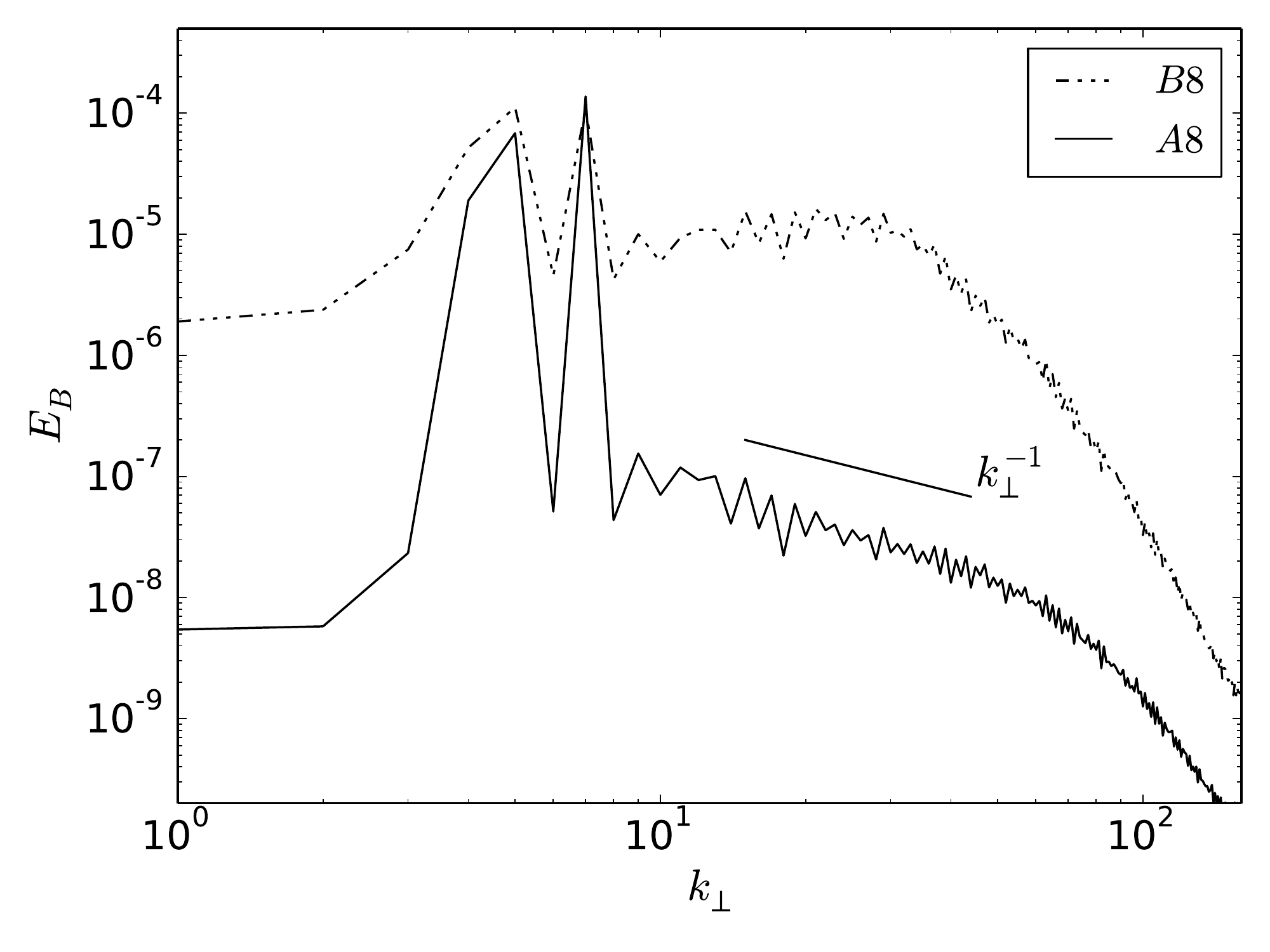}
\caption{{\it Top:} Power spectrum of magnetic field fluctuations for simulations with zero helicity (runs in set A). {\it Bottom:} Same for the two simulations with strongest guide field, runs A8 ($B_{0}=8$, $\alpha_h=0$)  and B8 ($B_{0}=8$, $\alpha_h=\pi/4$). A slope $k_\perp^{-1}$ is shown as a reference.}
\label{fig:Eb_vs_k}
\end{figure}

Strictly speaking these are not fluxes, as the kinetic energy, the kinetic helicity, and the magnetic energy are not conserved quantities in the ideal MHD limit. The flux of total energy $\Pi(k) = \Pi_v(k) + \Pi_b(k)$ is a flux, as the total (kinetic plus magnetic) energy is an ideal invariant of the MHD equations. As a result, $\Pi(k) \to 0$ for $k \to \infty$, and in the numerical simulations $\Pi(k_{max})=0$ with $k_{max}$ the maximum resolved wave number \cite{frisch_turbulence:_1995}. However, we still can consider the separate fluxes $\Pi_v(k)$ and $\Pi_b(k)$, and interpret them respectively as the fluxes of the kinetic and magnetic energy, plus the exchange of energy (i.e., work) done between the two fields \cite{mininni2005dynamo}. The same happens with the flux of kinetic helicity, which neglects all magnetic terms in the momentum equation, but which can represent a flux if magnetic fluctuations become negligible. Moreover, just as with the spectra, we can integrate any of these quantities over spheres to get isotropic fluxes $\Pi(k)$, or over cylinders in Fourier space to get perpendicular fluxes $\Pi(k_\perp)$.

In Fig.~\ref{fig:esquema_flujo} we show a diagram of how a typical flux (of kinetic energy, magnetic energy, or kinetic helicity) looks like in a simulation with moderate $B_0$. We define $\Pi^{+}$ as the maximum value of direct flux, $\Pi^{-}$ as the maximum value of inverse flux (i.e., the absolute value of the minimum of negative flux), and $\Delta \Pi$ as the value of the flux at $k=k_{max}$. As mentioned above, for an invariant quantity undergoing a cascade $\Delta \Pi$ should be zero. Indeed, $\Delta \Pi_v + \Delta \Pi_b = 0$ in all simulations, as the total energy is an ideal invariant which has a direct cascade in MHD turbulence. It follows that $\Delta \Pi_v = -\Delta \Pi_b$, which expresses the fact that the second terms on the r.h.s.~of Eqs.~(\ref{eq:PI_e_cin}) and (\ref{eq:PI_e_mag}) are associated with the exchange of energy between the magnetic and the velocity fields, which conserve the total energy when both energy components are added together. However, in the simulations magnetic field fluctuations ${\bf b}$ become negligible as $B_0$ is increased (see Fig.~\ref{fig:fluc_mag_b0}). In this case, the second term on the r.h.s.~of Eq.~(\ref{eq:PI_e_cin}) and both terms on the r.h.s.~of Eq.~(\ref{eq:PI_e_mag}) become negligible, and $\Delta \Pi_v$ can approach zero. If this happens, then the kinetic energy can be interpreted as a quantity conserved by nonlinear interactions in the inertial ranges (i.e., as a quantity that can have a cascade); the same argument applies to the kinetic helicity.

\begin{figure}
\centering
\includegraphics[width=8.3cm]{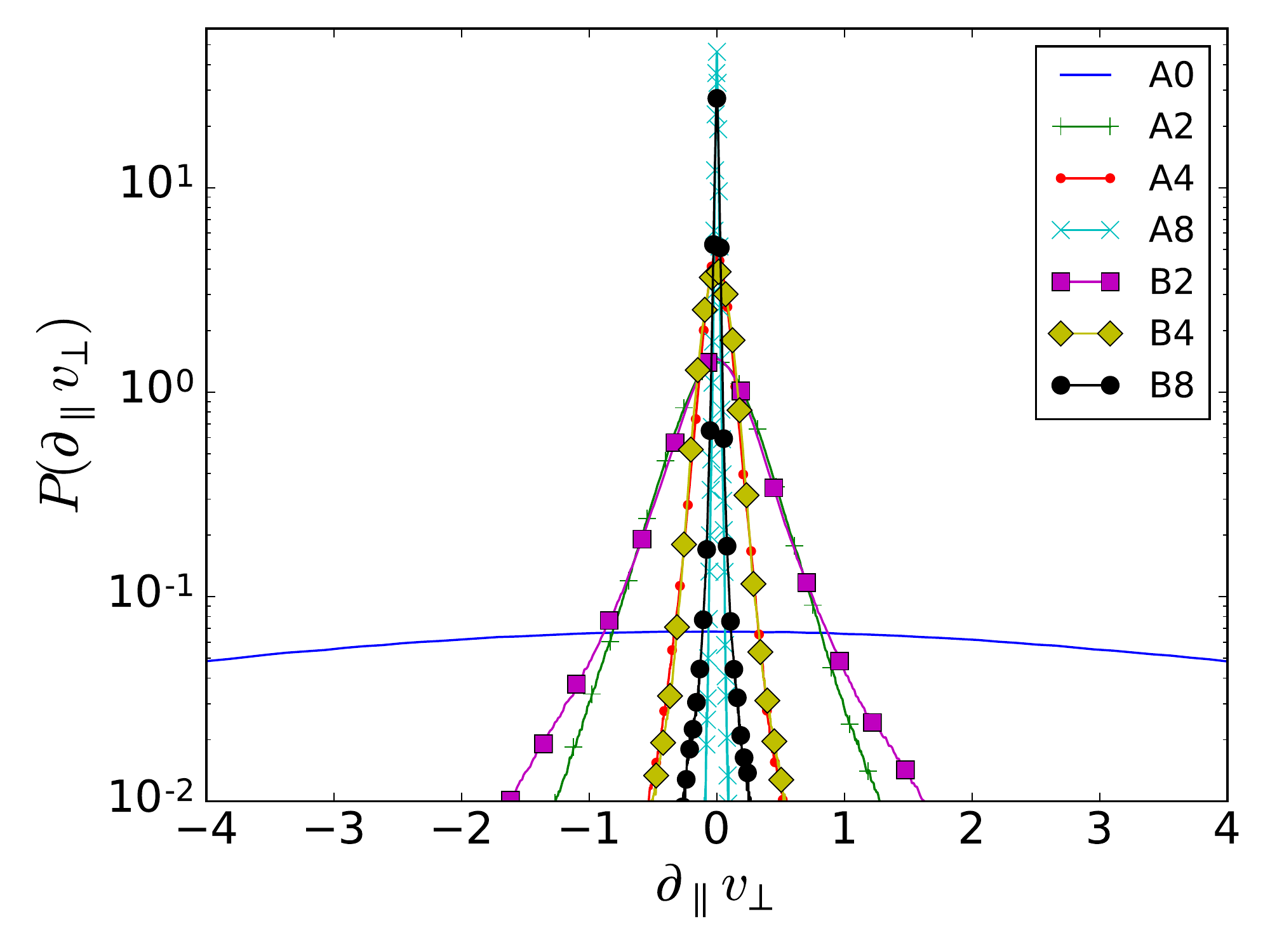}\\
\includegraphics[width=8.3cm]{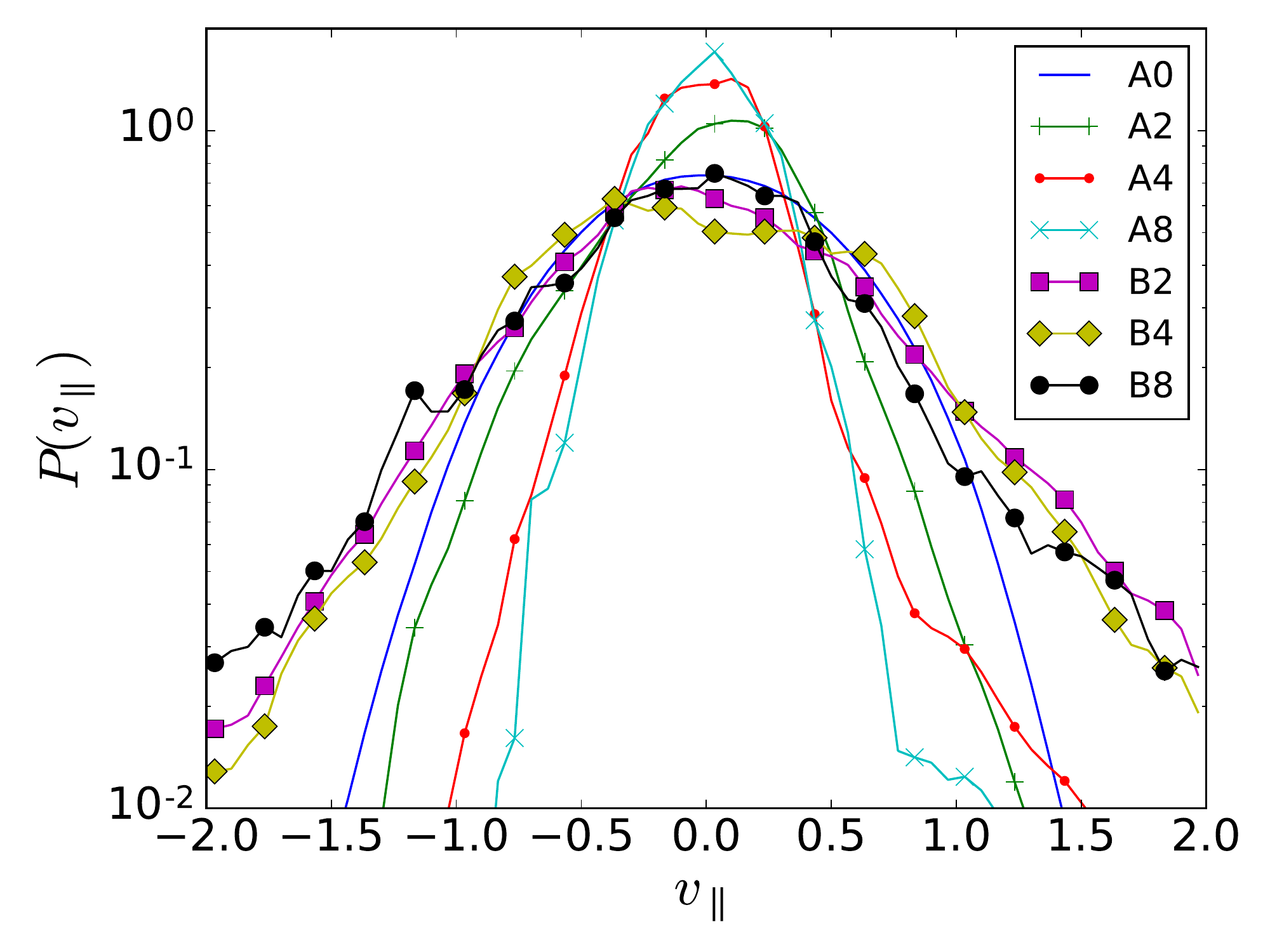}
\caption{{\it (Color online)} {\it Top:} Probability density functions (PDFs) of the parallel spatial derivative (i.e., the spatial derivative in the direction of the guide field) of a component of the velocity perpendicular to the guide field ($\partial_\parallel v_\perp$), for all runs. {\it Bottom:} PDFs of the component of the velocity parallel to the guide field ($v_\parallel$) for all runs.}
\label{fig:histogramas}
\end{figure}

The time-averaged kinetic energy fluxes for runs in sets A and B are shown in Fig.~\ref{fig:PIv_vs_k}. Indeed, $\Delta \Pi_v$ is almost negligible in the simulations with $B_0=4$ and $8$ (runs A4, A8, B4, and B8). This indicates that the system approaches a hydrodynamic regime for large $B_0$, independently of the helicity content of the flow. In this limit, the function $\Pi_v$ is indeed a flux. Moreover, in both sets of runs it is observed that $\Pi_v^{+}$ decreases and $\Pi_v^{-}$ increases with $B_0$. In other words, increasing the intensity of the field results in a suppression of the direct transfer of kinetic energy (compatible with the steeper energy spectrum observed in Fig.~\ref{fig:Ev_vs_k}), and in the development and increase of an inverse transfer (compatible with the growth of energy at small $k_\perp$ in Fig.~\ref{fig:Ev_vs_k}).

The kinetic helicity fluxes in runs in set B behave in a similar way, but with two notable differences. As in the case of the kinetic energy, $\Delta \Pi_h$ becomes negligible for large $B_0$. However, $\Pi_h^-$ does not increase with $B_0$ and instead it fluctuates around zero (as expected for a system without an inverse transfer of helicity), and as a result  $\Pi_h^+$ does not decrease as abruptly with $B_0$ as does $\Pi_v^{+}$. Figure \ref{fig:PIh_sobre_PIv} shows the ratio of the direct helicity flux $\Pi_{h}^{+}$ to the direct energy flux $\Pi_{v}^{+}$, as a function of $B_0$ for all the runs with helicity. To have a dimensionless ratio, and considering the Schwarz inequality, the direct helicity flux $\Pi_{h}^{+}$ is normalized by the helicity (and energy) injection wavenumber $k_f$, such that the ratio $\Pi_{h}^{+}/(k_f \Pi_{v}^{+})$ is unity when the kinetic energy and kinetic helicity fluxes are balanced. As $B_0$ increases, so does $\Pi_{h}^{+}/(k_f \Pi_{v}^{+})$. In other words, for large $B_{0}$ and with mechanical helical forcing, the direct cascade of helicity dominates over the direct transfer of kinetic energy to small scales.

\subsection{Scaling of helicity at small scales}

The results above indicate that in runs with injection of kinetic helicity and with a strong guide field, the inverse transfer of kinetic energy results in a diminished transfer of kinetic energy towards small scales. As a result, kinetic helicity, which can only be transferred towards smaller scales and which suffers a cascade in the HD limit, dominates the direct cascade. This is more clear in run B8, in which the normalized direct kinetic helicity flux is twice larger than the direct energy flux. This allows us to derive scaling laws for the kinetic energy and helicity spectra.

Let's assume that for large enough $B_0$, the direct flux of kinetic helicity is large enough that the direct flux of energy can be neglected. Moreover, as magnetic fluctuations are very small and the system is almost in a hydrodynamic regime, we can then assume that at the direct inertial range (i.e., for wave numbers larger than $k_f$) the helicity flux is approximately constant
\begin{equation}
\label{eq:cascada_heli}
\Pi_h^{+} \sim \sigma \sim \dfrac{\delta  h_\ell}{\tau_\ell} \dfrac{\tau_{a}}{\tau_\ell},
\end{equation}
where $\sigma$ is the helicity injection rate (equal to the helicity dissipation rate in the turbulent steady state),  $\delta h_\ell$ is the helicity at scale $\ell$, $u_\ell$ is the characteristic velocity of eddies of size $\ell$, $\tau_\ell \sim u_\ell/\ell_\perp$ is the eddy turnover time ($\ell_\perp$ is the eddy size in the direction perpendicular to ${\bf B}_{0}$, as the the eddies are almost 2D), and $ \tau_{a} \sim \ell_\parallel /B_{0} \sim L /B_{0}$ is the Alfv\'{e}n time (for large $B_0$, the characteristic length in the direction parallel to the guide field is the box size, i.e., $\ell_\parallel \sim L$). In isotropic and homogeneous turbulence, the helicity cascade rate (and the flux) would be estimated following Kolmogorov phenomenology as $\sigma \sim \delta  h_\ell/\tau_\ell$ (see, e.g., \cite{chen2003joint}). However, in the presence of Alfv\'en waves, the waves are expected to slow down the transfer linearly as the ratio of the two relevant time scales in the system (the Alfv\'en time and the turnover time) \cite{kraichnan1958irreversible,davidson_turbulence_2013}. Thus, the cascade rate for helicity in Eq.~(\ref{eq:cascada_heli}) must include the factor $\tau_{a}/\tau_\ell$.
%
%
Considering $E(k_\perp) \sim u_\ell^{2}/k_\perp$ and $H(k_\perp) \sim \delta h_\ell / k_\perp$, from Eq.~(\ref{eq:cascada_heli}) we obtain
\begin{equation}
\label{eq:E_H}
E(k_\perp)H(k_\perp) \sim \dfrac{1}{k_\perp^4}\dfrac{\sigma B_{0}}{L}.
\end{equation} 
Assuming $H(k_\perp) \sim k_\perp^{-h}$ and $E(k_\perp) \sim k_\perp^{-e}$ we obtain
\begin{equation}
\label{eq:sistema_exponentes}
e+h=4; \ \ \ h \geq e-1,
\end{equation}
where the first expression comes from Eq.~(\ref{eq:E_H}), and the second comes from the Schwartz inequality for $E$ and $H$. The equality holds for a flow with maximal helicity, in which case $e=5/2$ and $ h = 3/2$ (see the slopes shown as references in Fig.~\ref{fig:Ev_vs_k}). Note that in practice a turbulent system with maximal helicity cannot be obtained even with maximal helical forcing, as the development of instabilities and the growth of nonlinearities in the flow requires the system to depart from the state of maximal helicity (which makes the nonlinear terms exactly zero in the HD case) \cite{kraichnan1973helical}. 

It is interesting to note that similar scalings were predicted and observed in other systems that develop an inverse cascade of energy, and in which the helicity could then dominate the direct cascade range. Examples include the case of helical rotating turbulence \cite{pouquet10,mininni_rotating_2010}, and truncated versions of the Navier-Stokes equation \cite{Biferale13}. To see if the relation given by Eq.~(\ref{eq:sistema_exponentes}) is compatible with the data, we show in Fig.~\ref{fig:E_H_compensados} the product of the kinetic energy and helicity spectra compensated by $k_\perp^{-4}$ for run B8. We also show in this figure the kinetic helicity spectrum $H(k_\perp)$, compensated by $k_\perp^{-3/2}$ for the same run. If the spectra follow the predicted power laws, when compensated they should be flat in the inertial range. Indeed, both spectra show a reasonable agreement with the phenomenological argument and with Eq.~(\ref{eq:sistema_exponentes}).

From Eq. ~(\ref{eq:sistema_exponentes}) it also follows that the relative helicity should remain constant in the inertial range. The relative helicity is defined as
\begin{equation}
\label{eq:Hr}
h(k)=\dfrac{H(k)}{kE_{v}(k)},
\end{equation}
where $k$ can be replaced everywhere by $k_\perp$ in the anisotropic case. From Schwarz inequality, $h(k)$ and $h(k_\perp)$ can take values between $-1$ and $1$, with zero corresponding to the non-helical (i.e., mirror symmetric) case. In helical isotropic and homogeneous 3D HD turbulence, $h(k) \sim k^{-1}$ \cite{chen2003joint}. From Eq.~(\ref{eq:sistema_exponentes}), in the anisotropic case $h(k_\perp)$ should decrease slower than $k_\perp^{-1}$ if the direct cascade of kinetic helicity is dominant for wave numbers smaller than $k_f$. In fact, $h(k_\perp)$ should be independent of $k_\perp$ if the system is maximally helical.

Figure \ref{fig:Hr_vs_k} shows the relative helicity spectrum $h(k_\perp)$ for runs B4 and B8. Only in the dissipative range (i.e., for large perpendicular wave numbers) the relative helicity follows a $\sim k_\perp^{-1}$ decay, with a slower decrease for run B8. At intermediate wave numbers $h(k_\perp)$ varies slowly near $k=k_f=10$ (specially for run B8), and decreases slower than $\sim k_\perp^{-1}$ in the inertial range, in reasonable agreement with the phenomenological argument presented above.

\subsection{Energy dissipation rate}

The change in the fluxes and in the scaling laws followed by the kinetic energy at small scales when helicity is present should also have an impact in the energy dissipation rate of the system. Note that as magnetic field fluctuations are negligible for large $B_0$, most of the energy must dissipate as mechanical energy, whose rate of dissipation is given by $2\nu \Omega$, where
\begin{equation}
\Omega = \frac{1}{2} \int \omega^2 \; {\textrm d}V ,
\end{equation}
is the enstrophy. Figure \ref{fig:disip_b0} shows the mechanical energy dissipation rate as a function of $B_0$ for runs in sets A and B (i.e., respectively without and with helical mechanical forcing). For runs without helicity the energy dissipation rate decreases with increasing $B_{0}$, which is to be expected as the kinetic energy spectrum goes from a Kolmogorov spectrum (for $B_0 = 0$) to a steeper spectrum compatible with $\sim k_\perp^{-3}$, resulting in less excitation of fluctuations at small scales. However, for the simulations with helical forcing the energy dissipation rate either fluctuates or increases slowly with $B_0$. This is consistent with a shallower spectrum for the energy ($E_v \sim k_\perp^{-5/2}$ if helicity is maximal), and also indicates that a larger fraction of the energy is transferred to small scales in this case.

The inset in Fig. \ref{fig:disip_b0} also shows the kinetic energy dissipation rate normalized by the mechanical energy injection rate. This ratio is also important as the mechanical energy injection rate $\epsilon$ also depends on $B_0$. The ratio $2\nu \Omega/\epsilon$ varies only slowly with $B_0$, and increases as $B_0$ increases (i.e., $\epsilon$ behaves similarly as $2\nu \Omega$ does as $B_0$ is varied). As expected, the ratio goes towards a value close to unity for large values of $B_0$. This is to be expected as for strong guide fields the system is almost hydrodynamic (i.e., magnetic field fluctuations are negligible), and thus the energy injected in the system can only be dissipated through velocity field fluctuations. In other words, for the HD regime we expect the Ohmic dissipation to go to zero, and $\epsilon \approx 2\nu\Omega$ in the steady state (with the small difference $\epsilon - 2\nu\Omega = dE_v/dt$ being responsible for the slow growth of energy associated with the inverse cascade).

\subsection{Scaling of magnetic energy fluctuations}

From Table \ref{tab:datos_simulaciones}, we observe that the r.m.s.~magnetic fluctuations $ \left< | { \bf b } |^{ 2 } \right> ^{ 1/2 }_{ t } $ decrease as $B_0$ increases, being an order of magnitude less than the r.m.s.~velocity field fluctuations for $B_{0}=4$, and two orders of magnitude smaller for $B_{0} = 8$ (see also Fig.~\ref{fig:fluc_mag_b0}). Although magnetic field fluctuations are small for large $B_0$, it is still interesting to see how magnetic energy is distributed in different scales. Figure \ref{fig:Eb_vs_k} shows the energy spectrum of magnetic field fluctuations. As already mentioned, these fluctuations are created by the deformation of the guide field by the turbulent velocity field. This process of induction has already been observed in some experiments of MHD flows with a guide field using gallium \cite{bourgoin_magnetohydrodynamics_2002}. In this case, from dimensional analysis we can expect \cite{bourgoin_magnetohydrodynamics_2002}
\begin{equation}
\label{eq:E_b_spectrum}
E_{B}(k_\perp) \sim f B_{0}^{2} k_\perp^{-1}, 
\end{equation}   
where $f=f(U/B_{0},R_{e},R_{m})$ is a dimensionless factor. This power law is indicated in Fig.~\ref{fig:Eb_vs_k} as a reference. All spectra are in good agreement with the power law except for the runs with mechanical helicity injection, which depart from this law as $B_0$ increases. In Fig.~\ref{fig:Eb_vs_k} we show the behavior of the spectrum in run B8 (with helicity, and with $B_{0} = 8$), which shows the most dramatic departure with an almost flat spectrum $E_{B}(k_\perp)$. This indicates that small scale fluctuations of the velocity must be different in the helical and non-helical runs, as they are responsible for the deformation of the guide field and for the induction mechanism (see below).

\subsection{Velocity statistics and vertical gradients}

Finally, confirmation that the flows approach a 2D regime for large values of $B_0$ can be also obtained from field visualizations in real space, or from studying the statistical properties of the fields and of the field gradients in real space. In Fig.~\ref{fig:histogramas} we show the probability density function (PDF) of the velocity field gradient in the direction parallel to ${\bf B}_0$, of a component of the velocity field perpendicular to the guide field, i.e., $\partial_\parallel {\bf v}_\perp = {\bf B}_0/B_0 \cdot \nabla {\bf v}_\perp$. The PDF is very wide for run A0 (no guide field), and becomes narrower as $B_0$ is increased, indicating vertical gradients decrease with $B_0$ and confirming the transition of the flow towards a 2D regime for large $B_0$. However, the simulations with helical forcing (runs in set B) always show slightly stronger tails in the PDF than the simulations with non-helical forcing (runs in set A); compare, e.g., the PDFs of $\partial_\parallel {\bf v}_\perp$ for runs A8 and B8 in Fig.~\ref{fig:histogramas}.

Figure \ref{fig:histogramas} also shows the PDF of the component of the velocity field parallel to the guide field, $v_\parallel = {\bf v} \cdot {\bf B}_0/B_0$. Interestingly, the runs with helicity present a greater dispersion. This results from the combination of the direct transfer of kinetic helicity, and of the presence of the guide field which makes the flow quasi-2D. As the flow has to be helical at small scales, and as the vorticity is mostly aligned parallel to the guide field (resulting from the bidimensionalization of the flow), the flows in set B must keep larger values of the parallel velocity field (and correlated with the perpendicular velocity) to maintain the small scale helicity.

\section{Conclusions}

We studied the transition of a three-dimensional magnetohydrodynamic flow forced only mechanically as the strength of the guide field was increased. Two cases, one with non-helical mechanical forcing, the other with maximally helical mechanical forcing, were compared. The first case is similar to systems studied before by other authors \cite{alexakis_two-dimensional_2011}, in which a transition to a two-dimensional hydrodynamic regime was found, with properties reminiscent of those found in a phase transition \cite{seshasayanan_edge_2014,seshasayanan_critical_2016}, and with the strength of the guide field acting as the order parameter. The second case was not considered before, and although it shares similarities with the non-helical case, it also presents important differences.

In all cases the behavior of the system for large guide fields $B_0$ was found to be consistent with a transition towards a two-dimensional hydrodynamic regime. Magnetic field fluctuations become negligible (with r.m.s.~magnetic fluctuations decreasing as $b \sim B_0^{-2.2}$), velocity field fluctuations become anisotropic and dominate the total energy, and the kinetic energy spectrum grows at scales larger than the forcing scale. The development of an inverse transfer of kinetic energy was confirmed by the growth of a peak of the kinetic energy spectrum at the smallest available wave numbers in the domain, and by inspection of the kinetic energy flux which becomes negative at small wave numbers. In agreement with this behavior, simulations with non-helical forcing and large guide field show a small scale spectrum compatible with a power law $\sim k_\perp^{-3}$, which is the spectrum of energy in the direct cascade range of 2D HD turbulence, as already reported in \cite{alexakis_two-dimensional_2011}.

In the presence of mechanical helicity, the spectra at small scales (i.e., at wave numbers larger than the forcing wave number) change. For strong guide fields, the kinetic energy spectrum becomes shallower, and an even shallower spectrum of kinetic helicity develops. This is accompanied by a large transfer of helicity towards small scales, which dominates over the direct transfer of kinetic energy. In this case, the system seems to still evolve towards a quasi-two dimensional regime, but in which the three components of the velocity must be correlated (and non-negligible) to satisfy the constraint given by the amount of kinetic helicity in the flow. Thus, velocities along the direction of the guide field are larger than in the non-helical case, parallel velocity gradients (albeit still small) are also larger than in the former case, and the dissipation rate changes with the helical flows dissipating more kinetic energy than the non-helical ones.

Based on these results we presented a phenomenological argument that predicts a scaling for the kinetic energy and helicity spectra, respectively $E(k_\perp) \sim k_\perp^{-e}$ and $E(k_\perp) \sim k_\perp^{-h}$ with $e+h=4$ and $h \le e-1$ (with the equality holding in the maximally helical case), and which is in good agreement with the data. This scaling corresponds to a system in which the dynamics of the small scales are dominated by a direct cascade of kinetic helicity. Finally, while the small magnetic field fluctuations excited by induction follow a power law $\sim k_\perp^{-1}$ in the non-helical flow, in the helical case the changes in the small-scale velocity changes this scaling significantly.

There are several examples of different regimes of magnetohydrodynamic turbulence in the literature, and it is thus unclear whether a universal regime exists for which a unifying theory can be developed. The results presented here show another regime so far unexplored, in which the system behaves as a strongly anisotropic flow, in which energy self-organizes at large scales, and mechanical helicity is transferred towards small scales. Exploration of these different regimes can shed new light on the properties of turbulence in conducting fluids, relevant for space physics, industrial flows, and laboratory experiments.

\begin{acknowledgments}
The authors acknowledge support from grants PICT No. 2011-1529 and UBACYT 20020110200359. PDM acknowledges support from the Carrera del Investigador Cient\'{\i}fico of CONICET.
\end{acknowledgments}

\bibliography{ms}

\end{document}